\documentclass[12pt]{iopart}
\usepackage{graphicx}
\usepackage{subcaption}
\usepackage{xcolor}
\usepackage{hyperref}
\usepackage{siunitx}
\usepackage{comment}
\usepackage{cite}
\usepackage{lineno}
\usepackage[font=normalsize]{caption}

\usepackage{booktabs}
\usepackage{float}

\hypersetup{
    colorlinks=true,
    linkcolor=blue,
    filecolor=blue,      
    urlcolor=blue,
    citecolor=blue,
}

\begin{document}

\title{PINCH: Pipeline-Informed Noise Characterization in LIGO's Third Observing Run}

\author{Zach Yarbrough$^1$, Andre Guimaraes$^{1, 2}$, Prathamesh Joshi$^{3, 4, 5}$, Gabriela Gonz\'alez$^1$, Andrew Valentini$^6$}

\address{$^1$Louisiana State University, Baton Rouge, LA 70803, USA}
\address{$^2$ Michelin Inc., Greenville, SC 29605, USA}
\address{$^3$ Department of Physics, The Pennsylvania State University, University Park, PA 16802, USA}
\address{$^4$ Institute for Gravitation and the Cosmos, The Pennsylvania State University, University Park, PA 16802, USA}
\address{$^5$ School of Physics, Georgia Institute of Technology, Atlanta, GA 30332, USA}
\address{$^6$ Carthage College, Kenosha, WI 53140, USA}

\ead{zach.yarbrough@ligo.org}
\vspace{10pt}
\begin{indented}
\item[]
\end{indented}

\begin{abstract} We present a method to identify and categorize gravitational wave candidate triggers identified by matched filtering gravitational wave searches (pipelines) caused by transient noise (glitches) in gravitational wave detectors using Support Vector Machine (SVM) classifiers. Our approach involves training SVM models on  pipeline triggers which occur outside periods of excess noise to distinguish between triggers caused by random noise and those induced by glitches. This method is applied independently to the triggers produced by the GstLAL search pipeline on data from the LIGO Hanford and Livingston observatories during the second half of the O3 observing run. The trained SVM models assign scores to ambiguous triggers, quantifying their similarity to triggers caused by random fluctuations, with triggers with scores above a defined threshold being classified as glitch-induced. Analysis of these triggers reveals the distinct impact of different glitch classes on the search pipeline, including their distribution in relevant parameter spaces. We use metrics such as the Bhattacharyya coefficient and an over-representation ratio to quantify the consistency and prevalence of glitch impacts over time and across parameter spaces. Our findings indicate that some glitch types consistently produce triggers in specific regions of the parameter space, while others generate triggers that are more widely distributed. We observe that Scattered Light glitches appear differently in the search pipeline before and after a commissioning change, demonstrating how such detector changes appear in the pipeline’s response to certain glitch classes. This method provides a framework for understanding and mitigating the influence of non-Gaussian transients on gravitational wave search pipelines, with implications for improving detection sensitivity and better understanding noise populations.

\end{abstract}

%\tableofcontents
\newpage

\section{Introduction}
Gravitational wave observatories such as the Laser Interferometer Gravitational-Wave Observatory (LIGO) \cite{Aasi:2015aLIGO}, Virgo \cite{Acernese:2015Virgo}, and KAGRA \cite{Akutsu2021KAGRA} have achieved remarkable success in detecting hundreds of gravitational wave events since 2015 \cite{gwtc-3:2021vkt,gracedb}, ushering in a new era of multimessenger astrophysics. The sensitivity of these detectors is fundamentally limited by noise, which includes both stationary noise and transient, non-Gaussian noise events known as glitches \cite{aLIGO:2020wna}. Glitches arise from a variety of sources, both known and unknown \cite{LIGO:2021ppb,LIGO:2024kkz, Virgo:2022ysc, KAGRA:2020agh}, and can mimic or obscure astrophysical signals, significantly impacting search pipelines and their ability to discern true gravitational wave events \cite{LIGOScientific:2017tza, Cabero:2019orq}. 

Gravitational wave search pipelines are designed to identify astrophysical signals buried in detector noise, typically using matched filtering to extract candidate events from the data. These pipelines primarily target compact binary coalescences (CBCs), which arise from the merger of binary neutron star (BNS), neutron star–black hole (NSBH), and binary black hole (BBH) systems. These pipelines apply signal-based classification metrics to separate true signals from noise artifacts, such as tests that evaluate consistency with expected waveform morphology or seek coincidence across multiple detectors. Because glitches occur alongside real signals in the data, pipelines must unavoidably process both. To mitigate false alarms, pipelines rely on techniques that assess whether a candidate's time–frequency structure resembles that of a modeled waveform, and on requiring signal consistency across multiple observatories. Under the right conditions, however, glitches can masquerade as a genuine gravitational wave event, resulting in a false-positive detection \cite{LIGOScientific:2016gtq}. Additionally, the presence of glitches coincident with genuine signals can, in extreme cases, prevent detection entirely or obscure scientifically meaningful CBC parameters. This occurred in the case of GW170817, a BNS merger whose detection alert was delayed due to a glitch in one detector \cite{LIGOScientific:2016gtq,Canton:2013joa, Powell_2018,Pankow:2018qpo}. As the number and rate of confident gravitational-wave detections increase with improved detector sensitivity, such situations are likely to become more frequent. The constant interplay between signal and noise in LIGO data—and the ways in which search pipelines engage with both—offers valuable insight into the pipeline's response to different noise types, as well as the nature of the noise environment itself. 

To improve search robustness and to gain new insight into specific noise varieties, it is essential to understand how glitches appear from the perspective of the search pipeline. In this work, we present PINCH (Pipeline-Informed Noise CHaracterization), a systematic and broadly applicable method to identify and characterize search pipeline triggers that have been induced by glitches. PINCH is designed to be pipeline-agnostic and relies on Support Vector Machine (SVM) classifiers trained on triggers from periods of nominal detector behavior, enabling the separation of triggers arising from background fluctuations from those associated with transient noise.

We present results using triggers produced by the GstLAL pipeline \cite{Messick_2017,Ewing2024GstLAL,Hanna:2019ezx,Sachdev:2019vvd,Cannon_2012,Tsukada_2023} analyzing data LIGO Hanford (LHO) and LIGO Livingston (LLO) from the second half of LIGO's third observing run (O3b), which spanned from 11 November 2019, to 30 March 2020. While this work focuses on GstLAL, the method is pipeline-agnostic and will be applied to triggers from any matched-filter search pipeline in future studies. For internal analysis purposes, O3b is divided into week-long segments referred to as ``chunks,” with chunks 24 through 40 corresponding to this observing period. We examine how different glitch types impact the search pipeline by analyzing their prevalence and impact across key parameter spaces, as well as how these effects evolve over time across chunks.

Prior efforts to characterize these noise populations have included analyses of how different glitch types can produce triggers with high signal-to-noise ratio (SNR) in pipelines such as PyCBC \cite{Davis:2020nyf}, studies that attempt to distinguish between high-mass gravitational wave signals and glitches \cite{Lopez:2024xby}, works that model the population properties of glitch-induced triggers \cite{Ashton:2021tvz, Heinzel:2023vkq}, and work that projects where glitches would appear in CBC parameter spaces using glitch waveform models \cite{Bondarescu:2023jcx}. Our work builds on these efforts by using the response of a production-level search pipeline to real glitches in interferometer strain data, and presents the results in novel ways. Our approach offers a new, pipeline-informed perspective on transient noise. By investigating how existing types of glitches manifest in the parameter space of matched-filter pipelines, PINCH reveals how different noise types are ``seen'' by the pipeline. We demonstrate a method to leverage the pipeline as a tool to better understand glitches themselves, revealing the unique fingerprint of each glitch type in the eyes of the pipeline, and identifying which glitch types are most prevalent or disruptive. In doing so, PINCH helps to bridge the gap between search performance and detector characterization, enabling strategies for improvement that are mutually informed and broadly applicable across current and future pipelines.

\section{Background}

\subsection{Transient Noise}
\label{sec:noise}
Short-duration non-Gaussian instances of transient noise are commonly referred to as glitches. Glitches can negatively impact detector performance and search sensitivity, and are therefore regularly studied and characterized with the goal of eventual elimination \cite{LIGO:2024kkz}. Instances of excess power are identified via Omicron, a signal processing tool that uses Q-transforms to decompose data into different frequency bands to identify transients \cite{Robinet:2020lbf}. Omicron assigns glitches an SNR value, central frequency, timestamp, and other characterizing parameters. Glitches are visualized via spectrograms, a process that displays the relative power of each frequency that makes up the glitch over time in a way that makes the glitch's appearance or morphology distinguishable in a 2D image.

\begin{figure}[ht] 
    \hspace{0.1cm}
    \includegraphics[width=0.9\textwidth]{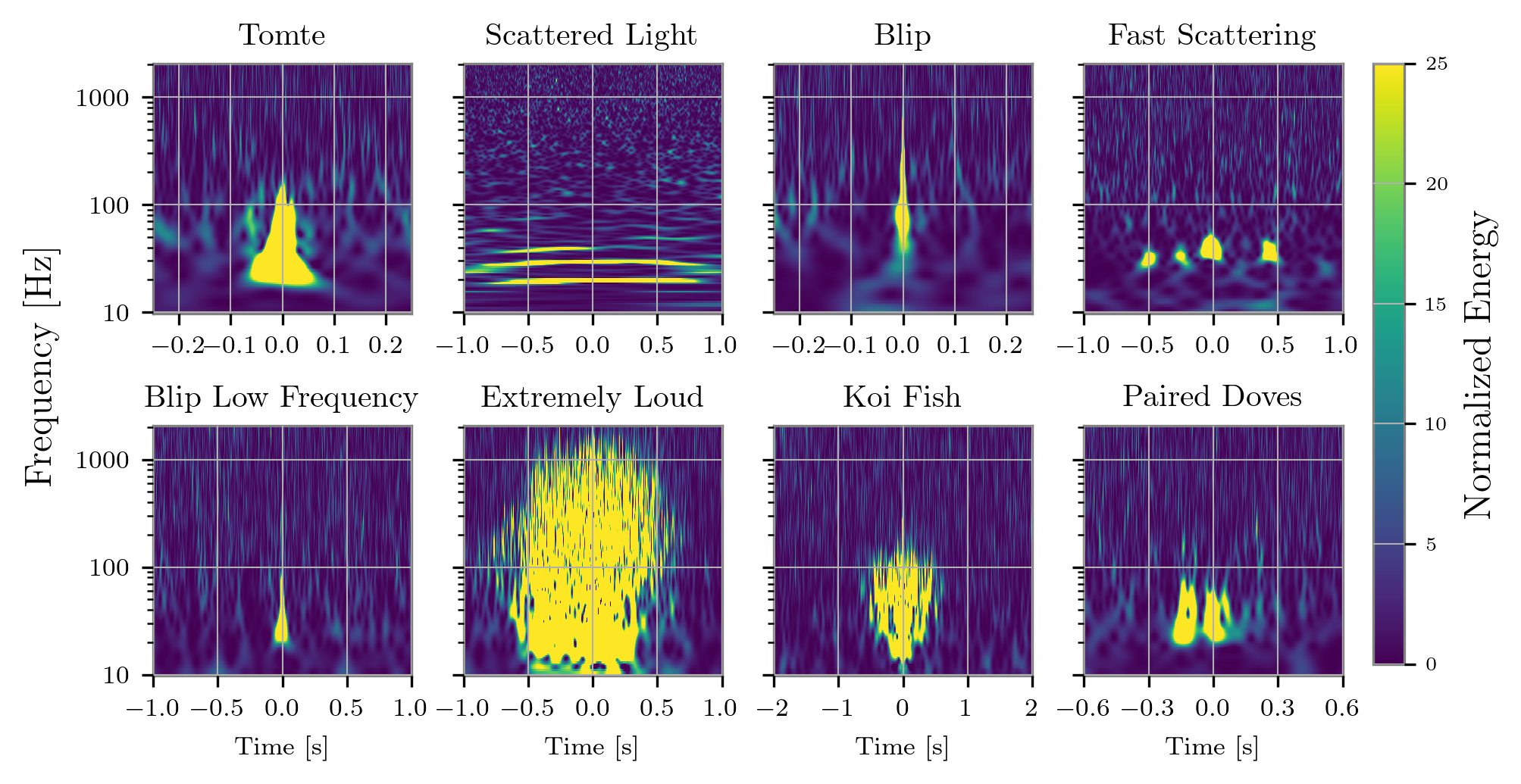}
    \caption{Spectrogram of several relevant glitch classes.}
    \label{fig:spectrogram}
\end{figure}

Glitches come in many shapes and sizes due to the variety of their origins. It is therefore useful to sort glitches as they occur into unique categories of similar morphologies. Doing so allows scientists to understand how specific types of glitches are impacting the detectors, and to make commissioning decisions that target specific glitch types. To achieve this categorization, a machine learning image classifier called Gravity Spy is used \cite{Glanzer:2022avx, Bahaadini:2018git, Soni:2021cjy, Mukund:2016thr, Wu:2024tpr}. Gravity Spy classifies glitches by comparing their spectrogram morphology against 23 different known morphologies. These categories include glitch types whose mechanisms are understood but sources are still unknown (such as Scattered Light, which is a repetitive, arch shaped glitch that is caused by errant light reentering the beam path after being reflected off of a surface), types whose mechanism and source are known but difficult to prevent (such as Whistles), and types whose source and mechanism are still unknown (such as Tomtes, a loud, broadband glitch that disproportionately impacts LLO). Spectrograms of the glitch types discussed in this work can be seen in Figure \ref{fig:spectrogram}.

Just as the unique Gravity Spy classes are useful for detector operations, it is interesting to analyze the impact of different glitch classes on the search pipeline. This is a new analysis that is made possible by the method outlined in this work. We show that different types of transient noise have unique signatures in the parameter spaces relevant to the GstLAL search pipeline.

\subsection{The GstLAL Search Pipeline}
\label{sec:gstlal}
GstLAL is a matched filtering pipeline based on a stream architecture that is designed to promptly detect signals from the merger of compact objects such as black holes and neutron stars \cite{Messick_2017,Ewing2024GstLAL,Hanna:2019ezx,Sachdev:2019vvd,Cannon_2012}. The matched filtering process compares thousands of pre-computed CBC waveform templates against observed detector data, producing a SNR timeseries that is discretized into events referred to as a triggers when the SNR crosses a threshold of 4 \cite{Messick_2017,Cannon_2012}. A signal-consistency test is evaluated for each trigger in order to determine how well the observed data agrees with the template across time that was used to produce the trigger \cite{Messick_2017}. A value for each of the component masses of the CBC system is associated with each trigger based on the masses used to compute the template waveform.

The GstLAL pipeline uses the matched-filter SNR $\rho$ and the quantification of the signal-consistency test $\chi^2$ of each trigger to construct a statistical noise background for each interferometer. This background is used to assess trigger significance and differentiate astrophysical signals from noise fluctuations. A detailed explanation of the GstLAL background construction can be found in \cite{Joshi:2023ltf}, but a brief summary is provided here. When a trigger occurs in one interferometer during a time when multiple detectors are simultaneously observing, the pipeline checks for a coincident trigger in another interferometer within a time window defined by the maximum light travel time between sites. If no such coincidence is found, the trigger is treated as noise and is added to a two-dimensional histogram built from the trigger’s $\rho$ and reduced $\chi^2$ values. This histogram, commonly referred to as the noise background or background, is constructed independently for each interferometer, and is used to estimate the probability density function \(P(\rho,  \chi^2 \mid \mathrm{noise})\) which is used in the GstLAL likelihood ratio (LR) statistic \cite{Tsukada_2023,Cannon:2012zt,2015arXiv150404632C}. Once the histograms have accumulated enough statistics, they are used to evaluate new triggers by comparing their $\rho$ and $\chi^2$ values to the empirical distribution of past noise. The likelihood ratio is computed for each candidate that meets the criteria to be considered by the pipeline instead of being immediately added to the background, and quantifies how signal-like a trigger is by comparing the probability of observing its parameters under a signal model versus under the noise background. The LR of these candidates is subsequently converted into a false alarm rate (FAR), which quantifies the expected rate at which noise alone would produce a trigger with equal or greater significance. FAR is typically expressed in units of time, such as events per year. Candidates with sufficiently low FARs are identified by the pipeline as confident gravitational wave events.

In this work, we examine the impact of transient noise on the GstLAL pipeline by analyzing the mass, $\rho$, and $\chi^2$ parameters of foreground triggers that are induced by glitches (all information about background triggers is lost once they are added to the background histogram). The overwhelming majority of these glitch-induced foreground triggers do not have sufficiently high likelihood ratios to be considered genuine gravitational wave detections. Nevertheless, by analyzing foreground triggers in the context of the glitch types that produce them, we are able to better understand the nature of transient noise using the pipeline, and simultaneously gain insight into how the pipeline perceives and responds to different glitch classes.

\section{Methods}

\subsection{Support Vector Machine Classifier}
To identify pipeline triggers caused by glitches, we construct an SVM model. An SVM is a supervised learning algorithm that seeks to separate data into two categories---in this case, triggers from nominal detector behavior versus glitch-induced triggers---by constructing an optimal decision boundary known as a hyperplane. This hyperplane maximizes the margin between the two classes in the chosen feature space. The model is trained on pipeline triggers that do not overlap with any periods of excess power, under the assumption that these represent nominal detector behavior. The process for identifying the triggers used for training is described below in \ref{sec:clean_dirty}. We use the trained model to evaluate triggers that do overlap with known transient noise, in order to separate them into two categories: those that are likely caused by the glitch they temporally coincide with, and those that merely occur at the same time but are not causally related. While the GstLAL pipeline identifies candidate triggers and assigns them likelihood-ratio statistics based on their consistency with modeled signals, it does not label individual noise-like triggers according to their origin. All non-coincident triggers are added to the background distribution, and the information about which noise source---if any---caused a specific trigger is lost. PINCH complements this approach by identifying which triggers were likely caused by specific glitches, enabling more targeted analysis of the relationship between glitch types and pipeline response. We build an independent SVM model for each chunk of O3b, training separate models for the Hanford and Livingston interferometers using their respective offline GstLAL triggers. That is, in each CBC observing chunk, one model is trained exclusively on Hanford triggers and used to evaluate Hanford data, while a separate model is trained and applied to Livingston triggers. This per-interferometer, per-chunk approach ensures that the classifier captures the unique noise characteristics of each detector over time.

\subsubsection{Clean vs. Dirty Triggers}
\label{sec:clean_dirty}

\begin{figure}[ht]
    \centering
    \includegraphics[width=0.6\textwidth]{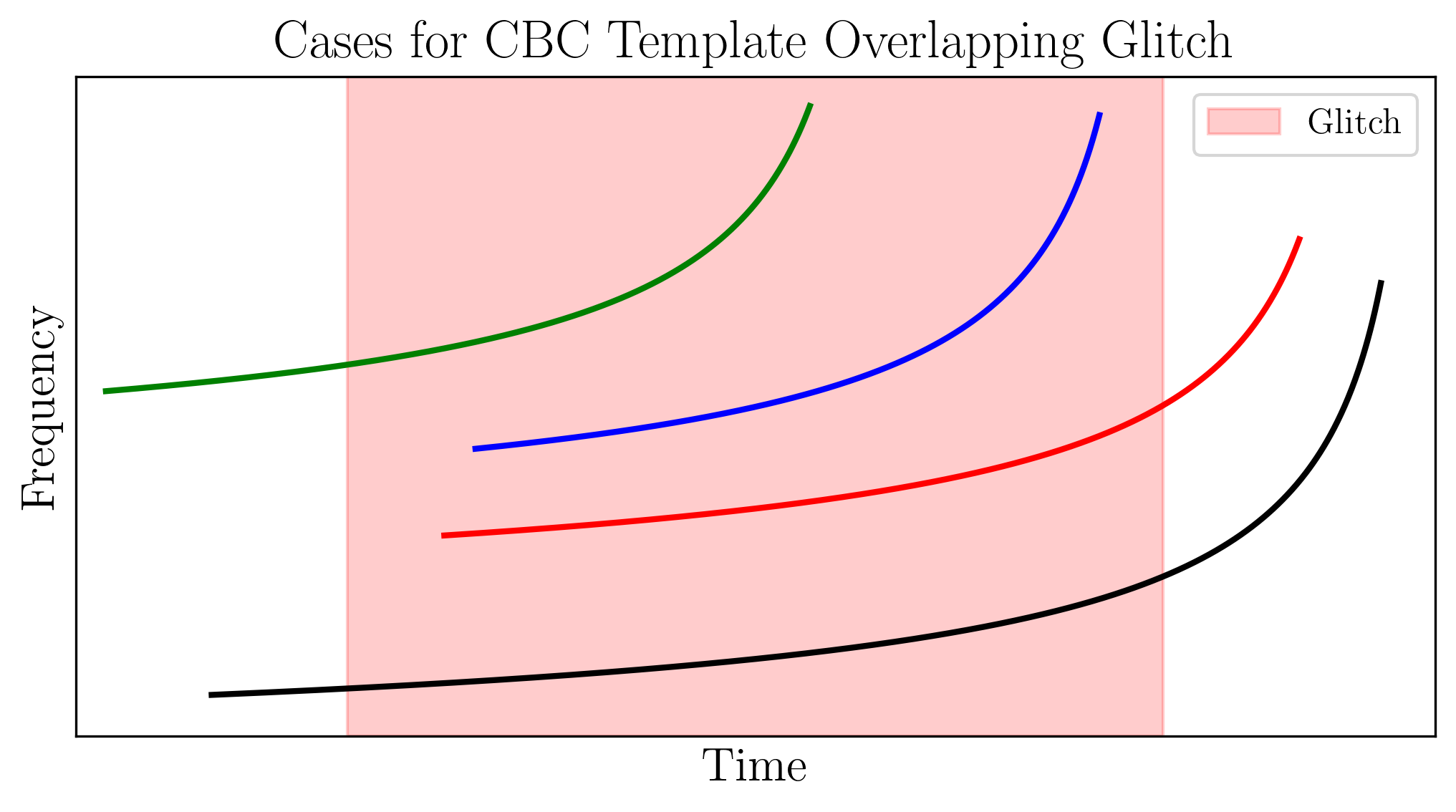}
    \caption{Possible cases for trigger template overlaps with a glitch of a given duration. Any pipeline trigger that meets one of these cases is labeled as dirty, and any trigger that starts and ends without overlapping with noise is called clean.}
    \label{fig:cases}
\end{figure}

Pipeline triggers are separated into two categories: clean and dirty. Clean triggers are pipeline triggers that do not occur during any times of transient noise, and are used as training data. Conversely, triggers are categorized as dirty if any transient noise occurs within a window preceding the trigger that begins at the start of the trigger's template duration and ends at the trigger's coalescence time. The width of this window is chosen because a trigger could be caused by noise if the noise occurs at any point in the trigger template duration. This categorization is possible by cross-referencing the coalescence time of the pipeline triggers with the times of known glitches produced by Omicron: if any Omicron trigger---times that are known to be transient noise---occurs within window of time defined by a trigger's template duration, that trigger is categorized as dirty. If no Omicron triggers appear in the window around a trigger, then it is considered clean. Since Omicron identifies times of excess power without distinguishing between instrumental artifacts and astrophysical signals, it may also produce triggers on genuine gravitational waves. Therefore, we exclude times of known gravitational wave events from this analysis. A visualization of this overlap is shown in Figure \ref{fig:cases}. The SVM is trained on the clean population to learn the characteristics of triggers that are unlikely to be caused by transient noise.  

An additional component of this categorization process is the association of different types of glitches to the triggers that they cause. When a pipeline trigger is labeled dirty due to the presence of an Omicron trigger, if that Omicron trigger has a large enough SNR to produce a Gravity Spy classification (as described in \ref{sec:noise}), the search pipeline trigger produced by this glitch is then associated with the particular glitch category that Gravity Spy identifies. Further categorizing the pipeline triggers caused by transient noise into their respective Gravity Spy classes allows us to understand how each class presents itself in the eyes of the search pipeline.

\subsubsection{Training}
\label{sec:training}

\begin{figure}[th]
    \centering
    \includegraphics[width=\textwidth]{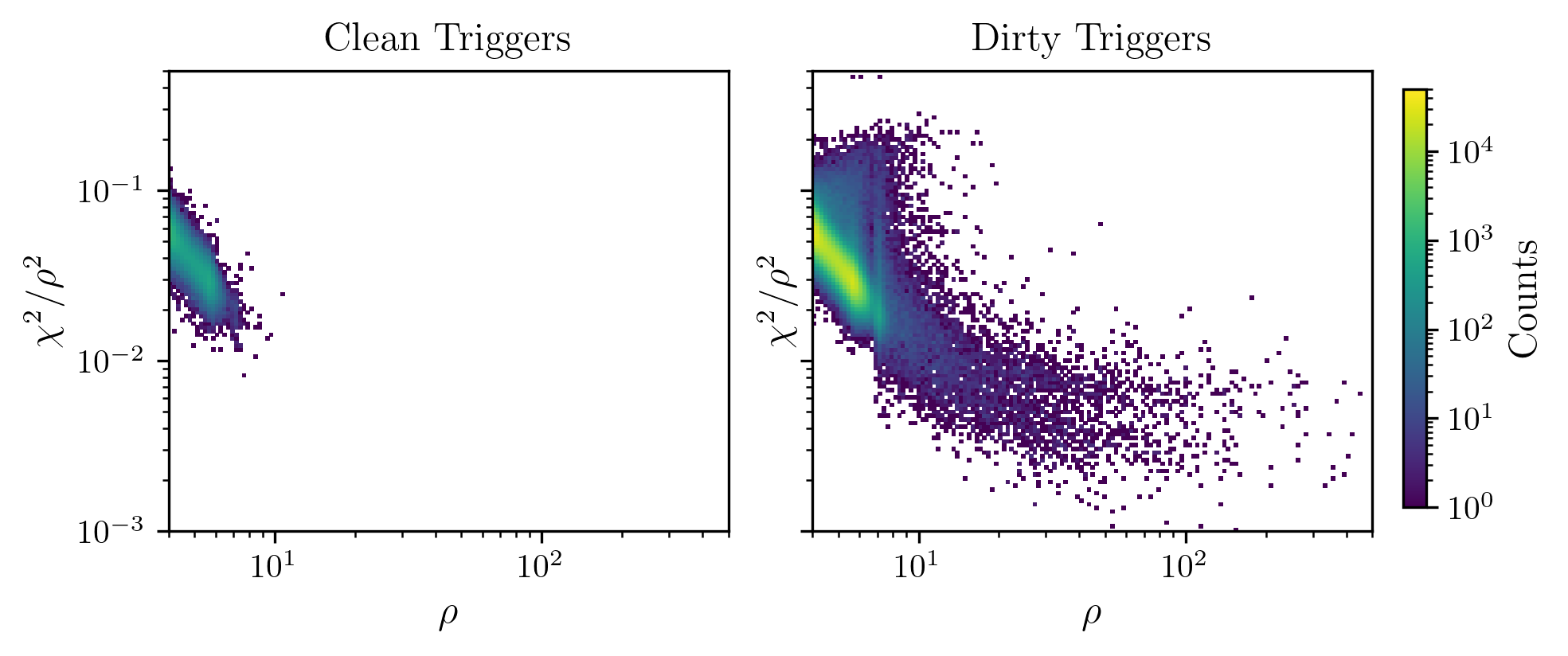}
    \caption{$\rho$ vs $\chi^2$/$\rho^2$ histograms for pipeline triggers at LLO in chunk 30 that do not overlap with transient noise and are labeled ``clean" (left) and those that do overlap with transient noise and have been labeled ``dirty" (right). Clean triggers are used as training data for the SVM, which is then used to evaluate the triggers in the dirty population.}
    \label{fig:dirty_v_clean}
\end{figure}

Once the pipeline triggers have been separated into clean and dirty groups, the SVM is trained on the clean set. A separate SVM model is created for each chunk of O3b, with distinct models trained for each interferometer. Each model is trained on the $\rho$ and $\chi^2$ values associated with the triggers, using 10,000 training samples per model. The result is an SVM capable of quantifying the similarity of a given trigger to the clean training distribution. Figure \ref{fig:dirty_v_clean} shows the difference in $\rho$ and $\chi^2$ between the clean training set and the dirty set that is scored by the trained SVM.

\begin{figure}
    \centering
    \includegraphics[width=\textwidth]{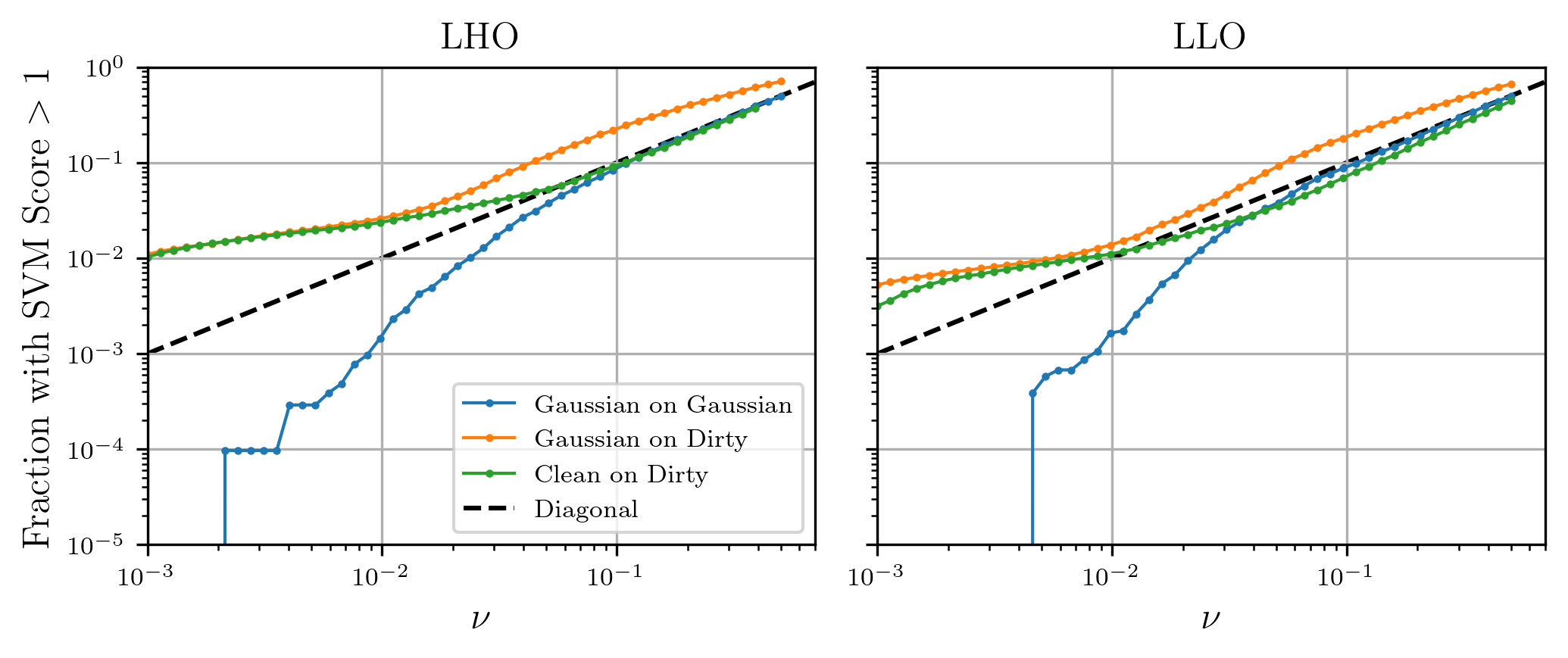}
    \caption{Fraction of evaluation data identified as outliers (SVM score $> 1$) versus training hyperparameter $\nu$ for various combinations of training and evaluation sets. At large $\nu$, all models approach the diagonal (dashed line), indicating the output is dominated by the training parameter rather than underlying data structure. Models trained with $\nu = 0.01$ return a more stable and selective set of outliers, motivating its use as a conservative threshold.}
    \label{fig:nu_svm}
\end{figure}

Evaluating the performance of the SVM model is challenging because the fraction of true outliers in the evaluation set is unknown, making standard validation techniques inapplicable. To address this, we sweep across a range of values for the SVM's hyperparameter $\nu$ to identify a conservative setting---one that may miss some true outliers but minimizes false positives. The value of $\nu$ ranges from $(0, 1]$ and defines the fraction of the training data that may be outliers, effectively relaxing the boundary between the two classes. It is referred to as a hyperparameter because it is set in advance and controls how the model is trained. To assess its effect on the fraction of outliers returned and inform the choice of a conservative setting, we trained SVM models across multiple $\nu$ values for several training and evaluation combinations. Figure \ref{fig:nu_svm} shows $\nu$ versus the fraction of evaluation data flagged as outliers (i.e., receiving an SVM score $> 1$).

When the model is trained on pipeline triggers produced by performing matched filtering on Gaussian noise and evaluated on a separate set of triggers produced in the same way, we expect to find a very small fraction of the triggers evaluated by the SVM to have a score $> 1$, since they are taken from the same set as the training triggers. We see this expected result at low $\nu$, however, at larger $\nu$ values, this curve converges toward the diagonal, indicating that the model increasingly returns exactly $\nu$-fraction of the data as outliers, independent of the true data structure. A similar trend is seen when models are trained on Gaussian or clean triggers and evaluated on the same set of pipeline triggers that overlap with glitches. These curves also approach the diagonal at large $\nu$, suggesting that high $\nu$ values are again uninformative and motivating the selection of a smaller $\nu$ value so as not to overestimate the number of outliers, rather than simply echoing the training parameter. Based on this behavior, we identify $\nu = 0.01$ as a practical training value at which the model returns a confident fraction of outliers by avoiding the diagonal regime where $\nu$ directly dictates the output.

While training models with a $\nu$ value of 0.01 technically results in more than 1\% of the evaluation set being flagged as unlike the training data, this level of over-counting relative to the training setting is still substantially more conservative than what occurs at higher $\nu$ values. The subset of outliers identified by models with $\nu = 0.01$ is more tightly constrained and thus reflects a higher confidence in the data identified by the models as being unlike their training data. Therefore, we selected $\nu = 0.01$ as a conservative value to reduce risk of over-counting and ensure the triggers identified by the SVM are confidently caused by glitches.

\subsubsection{Application of SVM}

\begin{figure}
    \centering
    \includegraphics[width=0.95\textwidth]{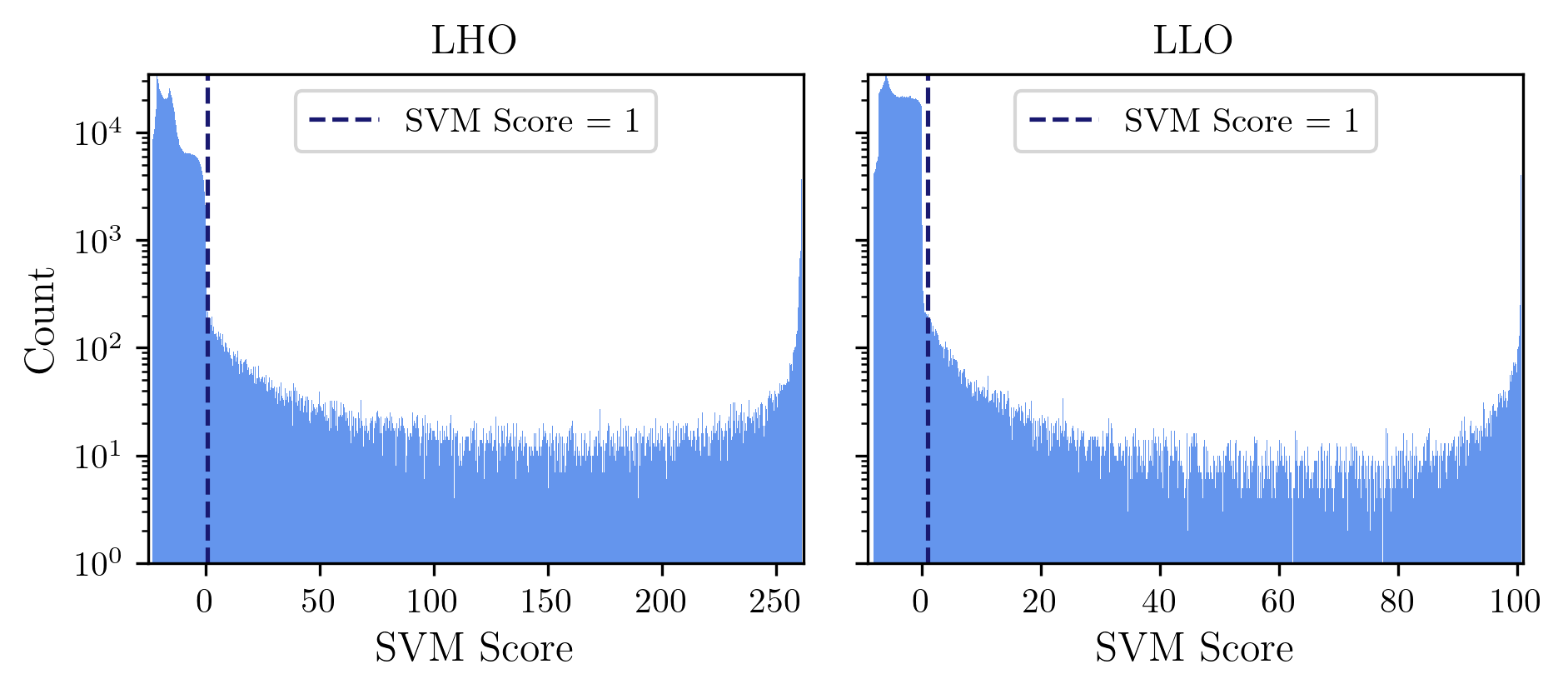}
    \caption{Distribution of scores produced by the SVM model when used to evaluate the similarity of the dirty triggers to the clean training data at LHO (left) and LLO (right) in the 30th chunk of O3b. The boundary at SVM score = 1 shows the confident boundary established by the SVM between triggers that are similar to the clean data and triggers that are not. Triggers with an SVM score $>$ 1 are likely to be caused by transient noise.}
    \label{fig:svm_distribution}
\end{figure}

After training the SVM on clean triggers, we use it to assign a score to each trigger labeled as dirty. This scoring is a quantization of the boundary that the SVM constructs between the two categories it is asked to differentiate between. Since the SVM is trained on clean triggers, it assigns a higher score to the triggers in the dirty dataset that are more similar to the clean training data, and a lower score to the triggers that are unlike the clean training data. By convention we invert the sign of the SVM scores so that the triggers more likely to be caused by glitches have higher SVM scores. The decision boundary for a binary SVM classifier like this one is nominally zero, however it is common to use a value of $\pm 1$ to ensure accurate categorization. Thus, the dirty triggers that are evaluated by the SVM and return a score greater than 1 are considered to be caused by glitches. Figure \ref{fig:svm_distribution} shows the distribution of SVM scores for the dirty triggers at each interferometer for the 30th chunk of O3, along with the decision boundary at SVM score equal to one. 

\begin{figure}[ht]
    \centering
    \includegraphics[width=0.95\textwidth]{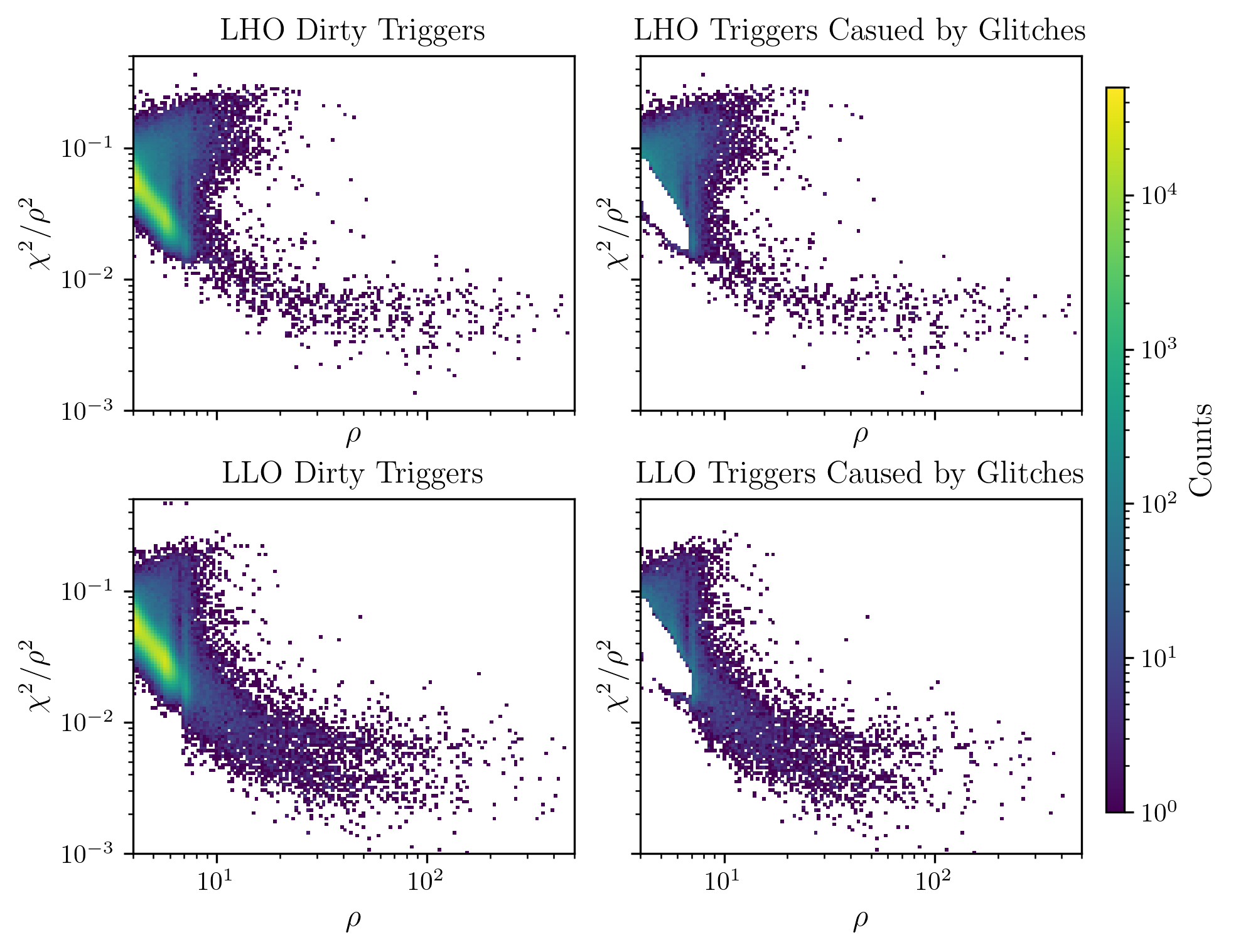}
    \caption{$\rho$ vs $\chi^2$/$\rho^2$ histograms for LHO (top) and LLO (bottom) showing the concentration of triggers categorized as dirty (left), and those which the SVM model determined to be caused by glitches (right) in the 30th chunk of O3b. The blank, ``cut out" region in the right column is the effect of removing triggers with SVM scores $\leq$ 1. The remaining triggers that populate the right column are those likely caused by glitches.}
    \label{fig:beforeafter}
\end{figure}

This process is repeated independently for each interferometer and for each chunk, where the triggers with an SVM score greater than 1 are retained for further analysis. Figure \ref{fig:beforeafter} shows all triggers at each interferometer in chunk 30 that have been labeled as dirty and the remaining triggers after the triggers with SVM scores less than 1 are removed.

This approach goes beyond simply removing low-SNR or high–$\chi^2$ triggers. In fact, many clean triggers fall within those regions of parameter space and are included in the training data. Rather than applying fixed thresholds, the SVM learns the multivariate structure of clean behavior in the joint SNR–$\chi^2$ space and identifies triggers that deviate from that distribution. This enables the model to flag glitch-induced triggers that might appear typical when viewed along any single axis but are statistically distinct when considered in combination. In this way, machine learning offers a flexible and adaptive method for distinguishing glitch-related triggers in a manner that complements existing pipeline techniques.

\subsection{Data Analysis Techniques}

After the SVM has determined which pipeline triggers are caused by glitches, we use this population to understand how specific types of noise manifests in the search pipeline. Here we discuss several tools that we use to understand the impact of various glitch types on the pipeline.

\subsubsection{Battacharyya Coefficient }
\label{sec:bc}

In order to determine whether the parameters of the triggers produced by a given glitch class are consistent or vary over time, we use the Bhattacharyya coefficient, a quantity that represents how similar two probability distributions are to one another \cite{bhattacharyya1943measure, duda2000pattern, bishop2006pattern}. For a discrete probability distribution, the Battacharyya coefficient is defined as 

\begin{equation}
BC(P, Q) = \sum_{x \in \chi} \sqrt{P(x)Q(x)}
\label{eq:bc}
\end{equation}

where $P(x)$ and $Q(x)$ are two discrete probability distributions over the same domain of parameter values, and $x$ indexes the binned quantity such as SNR or component mass. The varible $\chi$ denotes the common support of these distributions, i.e. the set of bins where both $P$ and $Q$ are defined (and is unrelated to the $\chi^2$ parameter discussed in the context of pipeline triggers). This metric returns a value between 0 and 1, with 0 representing no overlap between probability distributions, and 1 representing a complete overlap. In practice, we compute $P$ and $Q$ as normalized histograms of glitch-induced trigger parameters from two different observing chunks, allowing us to quantify the similarity of these distributions using the Bhattacharyya coefficient. Bhattacharyya coefficients for relevant distributions are shown in Figure \ref{fig:bc_matrix} and discussed in Section \ref{sec:variation}.

\subsubsection{Over-Representation Ratio}

In order to quantify the impact of different glitch types in the perspective of the pipeline, we introduce the over-representation ratio, defined as

\begin{equation}
   ORR = \frac{N_{PG}}{N_{\text{pipeline}}} \cdot\frac{N_{\text{glitch}}}{N_{\text{glitch type}}}
\label{eq:overrep}
\end{equation}
where $N_{\text{pipeline}}$ and $N_{\text{glitch}}$ are the total number of pipeline triggers and Gravity Spy glitches respectively over the given time interval, $N_{\text{glitch type}}$ is the number of glitches from a particular glitch class that occur over the same time interval, and $N_{PG}$ is the number of pipeline triggers over the same time interval that are caused by the glitch class in question. In this work, we compute the ORR using data from the full O3b observing period in order to provide a global assessment of the impact of each glitch type.

This quantity relates the percentage of pipeline triggers caused by each glitch class to the percentage of glitches that each glitch class constitutes in a single dimensionless value. A higher over-representation ratio is caused by a certain glitch type causing a large number of pipeline triggers relative to the number of occurrences of that particular variety of noise. Thus, a higher over-representation ratio indicates a more potent glitch in the context of the pipeline. Figure \ref{fig:overrep_combo} shows the over-representation ratio for a variety of glitch types for each interferometer.

\section{Results}

To demonstrate the results of applying PINCH to pipeline triggers triggers, we use data O3b . We use triggers from the GstLAL search pipeline running in offline mode on strain data from the LIGO Hanford and LIGO Livingston detectors. This stretch of data is broken into 17 smaller segments of time referred to as chunks. O3b consisted of chunks 23 through 40. As mentioned in Section \ref{sec:training}, we construct one SVM model for each interferometer per chunk in order to capture the unique variations in transient noise over time and between observatories.

\begin{table}
\centering
\begin{tabular}{c c c r r r r}
\toprule
 Chunk &    Start &      End &     Total & Clean \% & Glitched \% & Gravity Spy \% \\
\midrule
    23 & 01-11-19 & 10-11-19 & 1,716,227 &    8.1\% &    1.06\% &  54.9\% \\
    24 & 10-11-19 & 18-11-19 & 1,472,890 &    4.4\% &    2.27\% &  69.2\% \\
    25 & 18-11-19 & 25-11-19 & 1,535,689 &    7.2\% &    1.01\% &  57.0\% \\
    26 & 25-11-19 & 03-12-19 & 1,298,006 &    6.8\% &    2.47\% &  70.8\% \\
    27 & 03-12-19 & 11-12-19 & 1,557,935 &   10.5\% &    0.78\% &  46.4\% \\
    28 & 11-12-19 & 20-12-19 & 3,496,142 &    6.3\% &    4.53\% &  75.9\% \\
    29 & 20-12-19 & 28-12-19 & 1,874,027 &   11.2\% &    0.93\% &  56.4\% \\
    30 & 28-12-19 & 04-01-20 & 1,289,265 &   10.8\% &    2.57\% &  76.2\% \\
    31 & 04-01-20 & 13-01-20 & 1,575,744 &    6.7\% &    6.95\% &  75.5\% \\
    32 & 13-01-20 & 22-01-20 & 1,467,596 &   10.3\% &    1.43\% &  59.4\% \\
    33 & 22-01-20 & 31-01-20 & 2,982,420 &   11.6\% &    0.91\% &  44.9\% \\
    34 & 31-01-20 & 07-02-20 & 2,740,192 &   12.4\% &    0.78\% &  39.8\% \\
    35 & 07-02-20 & 14-02-20 & 1,368,536 &   13.2\% &    0.61\% &  38.9\% \\
    36 & 14-02-20 & 23-02-20 & 1,605,294 &   12.6\% &    0.68\% &  42.5\% \\
    37 & 23-02-20 & 02-03-20 & 1,531,571 &   12.2\% &    0.97\% &  46.2\% \\
    38 & 02-03-20 & 09-03-20 & 1,454,012 &   13.3\% &    0.61\% &  35.9\% \\
    39 & 09-03-20 & 16-03-20 & 1,484,364 &   13.2\% &    0.66\% &  37.8\% \\
    40 & 16-03-20 & 30-03-20 & 2,053,624 &   13.2\% &    0.81\% &  26.5\% \\
\bottomrule
\end{tabular}
\caption{Summary of LIGO Hanford pipeline triggers and glitch associations during O3b. The total number of pipeline triggers is shown in the Total column, followed by the percentage of pipeline triggers that are labeled clean because they do not overlap with transient noise. The Glitched column now shows the percentage of total triggers attributed to glitches based on SVM analysis. The Gravity Spy column is the percentage of the glitched triggers that are caused by a glitch with a gravity spy classification $\geq$90\%.}
\label{tab:nu_o3b_h1_summary_mod}
\end{table}

\begin{table}
\centering
\begin{tabular}{c c c r r r r}
\toprule
 Chunk &    Start &      End &     Total & Clean \% & Glitched \% & Gravity Spy \% \\
\midrule
    23 & 01-11-19 & 10-11-19 & 1,444,298 &    3.8\% &    2.06\% &  71.9\% \\
    24 & 10-11-19 & 18-11-19 & 1,580,279 &    0.9\% &    1.91\% &  73.6\% \\
    25 & 18-11-19 & 25-11-19 & 1,676,994 &    2.9\% &    1.91\% &  73.2\% \\
    26 & 25-11-19 & 03-12-19 & 1,401,329 &    2.4\% &    1.68\% &  69.0\% \\
    27 & 03-12-19 & 11-12-19 & 1,629,790 &    4.3\% &    1.71\% &  55.5\% \\
    28 & 11-12-19 & 20-12-19 & 3,152,062 &    1.7\% &    3.96\% &  74.4\% \\
    29 & 20-12-19 & 28-12-19 & 2,019,891 &    2.4\% &    2.59\% &  74.7\% \\
    30 & 28-12-19 & 04-01-20 & 1,790,641 &    3.4\% &    1.39\% &  68.0\% \\
    31 & 04-01-20 & 13-01-20 & 1,587,384 &    2.8\% &    1.47\% &  68.1\% \\
    32 & 13-01-20 & 22-01-20 & 1,340,965 &    2.0\% &    1.18\% &  64.2\% \\
    33 & 22-01-20 & 31-01-20 & 3,577,272 &    2.4\% &    1.96\% &  58.5\% \\
    34 & 31-01-20 & 07-02-20 & 2,996,950 &    3.4\% &    1.75\% &  61.4\% \\
    35 & 07-02-20 & 14-02-20 & 1,587,635 &    2.9\% &    1.44\% &  64.6\% \\
    36 & 14-02-20 & 23-02-20 & 1,535,792 &    3.7\% &    1.74\% &  62.9\% \\
    37 & 23-02-20 & 02-03-20 & 1,646,038 &    4.7\% &    1.50\% &  65.6\% \\
    38 & 02-03-20 & 09-03-20 & 1,593,177 &    4.7\% &    0.95\% &  54.3\% \\
    39 & 09-03-20 & 16-03-20 & 1,642,923 &    6.1\% &    1.16\% &  55.2\% \\
    40 & 16-03-20 & 30-03-20 & 2,274,056 &    5.3\% &    1.12\% &  34.7\% \\
\bottomrule
\end{tabular}
\caption{Summary of LIGO Livingston pipeline triggers and glitch associations during O3b. The total number of pipeline triggers is shown in the Total column, followed by the percentage of pipeline triggers that are labeled clean because they do not overlap with transient noise. The Glitched column now shows the percentage of total triggers attributed to glitches based on SVM analysis. The Gravity Spy column is the percentage of the glitched triggers that are caused by a glitch with a gravity spy classification $\geq$90\%.}
\label{tab:nu_o3b_l1_summary_mod}
\end{table}

To better understand the distribution of this categorization during O3b, we summarize the number of total pipeline triggers, triggers labeled as clean and dirty, and the subset of triggers identified as glitch-induced by the SVM, as well as those triggers induced by glitches with a Gravity Spy classification with confidence $\geq 90\%$. Tables \ref{tab:nu_o3b_h1_summary_mod} and \ref{tab:nu_o3b_l1_summary_mod} provide a breakdown for these quantities for each chunk of O3b at LHO and LLO, respectively. The number of pipeline triggers confidently caused by glitches is relatively small compared to the total number of triggers generated over each of the observing chunks. For LIGO Hanford, the chunk with the smallest number of glitch-induced triggers is chunk 35, with 8,348 such triggers out of 1.37 million total (0.61\%). The largest number of glitch-induced triggers occurs in chunk 28, with 158,375 out of 3.5 million (4.53\%). In terms of percentages, the chunks with the lowest fraction of pipeline triggers caused by glitches are chunks 35 and 38, with 0.61\%, while the highest is in chunk 31, where 6.95\% of the 1.58 million triggers were identified as glitch-induced.
At LLO, the fewest glitch-induced triggers are found in chunk 38, with 15,135 out of 1.59 million total (0.95\%), while the most occur in chunk 28, with 124,822 out of 3.15 million triggers (3.96\%). The lowest percentage of glitch-induced triggers is in chunk 38 (0.95\%), and the highest is in chunk 28 (3.96\%), reflecting both the relative and absolute prominence of glitches in that period. The majority of triggers in each chunk are neither labeled as clean nor confidently glitch-induced. These are triggers that coincide with transient noise but either lack a Gravity Spy classification with $\geq 90\%$ confidence or are not confidently categorized by the SVM. While the overall number of glitch-induced triggers remains low, the variation in relative percentages across chunks highlights variation in detector behavior and pipeline response.

\subsection{Pipeline Response to Transient Noise}

\subsubsection{O3b Overall}

\begin{figure}[ht]
    \centering
    \begin{subfigure}{0.89\textwidth}
        \centering
        \includegraphics[width=0.89\textwidth]{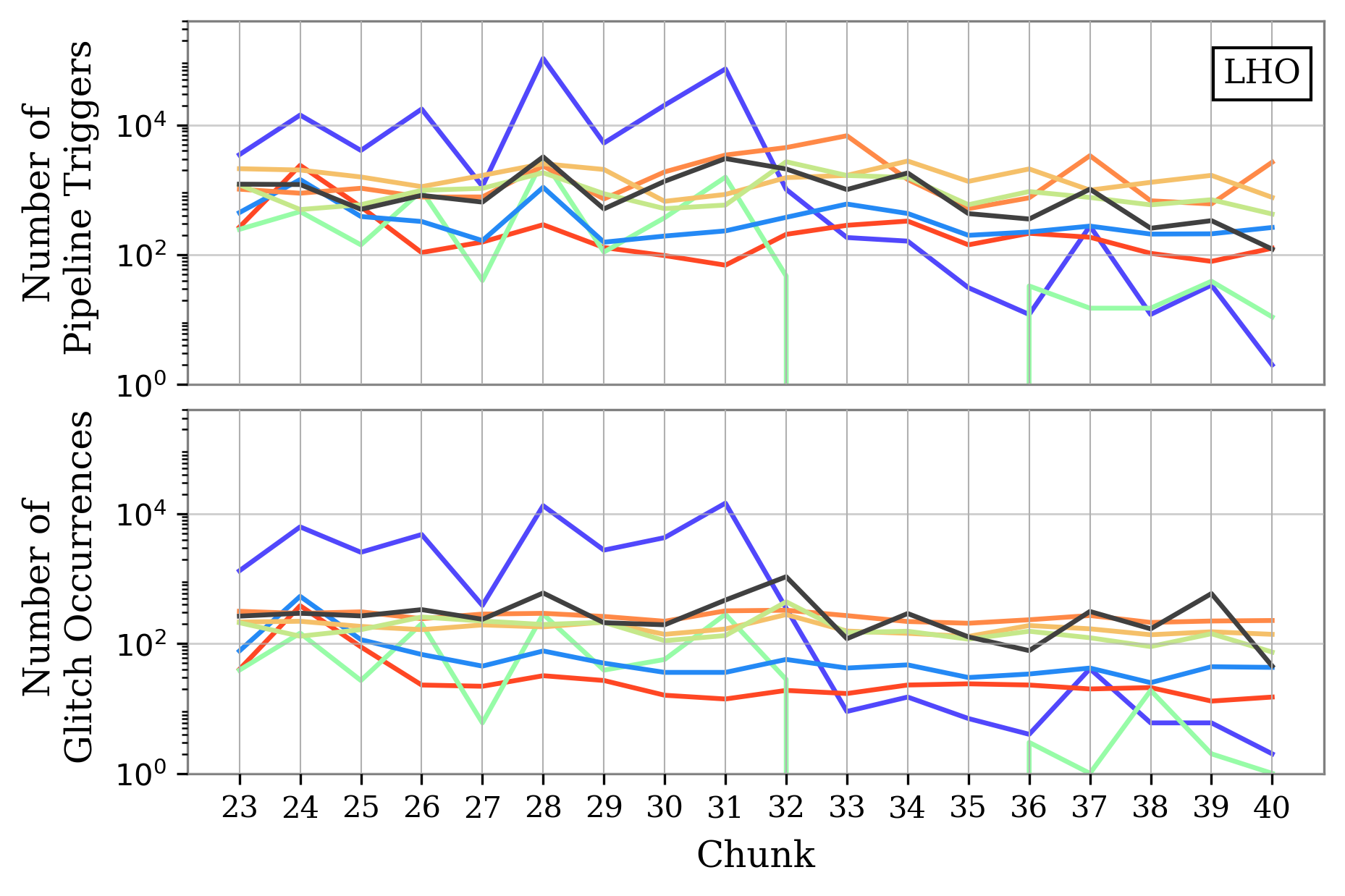}
    \end{subfigure}
        \begin{subfigure}{0.89\textwidth}
        \centering
        \hspace{0.1cm}
        \includegraphics[width=0.89\textwidth]{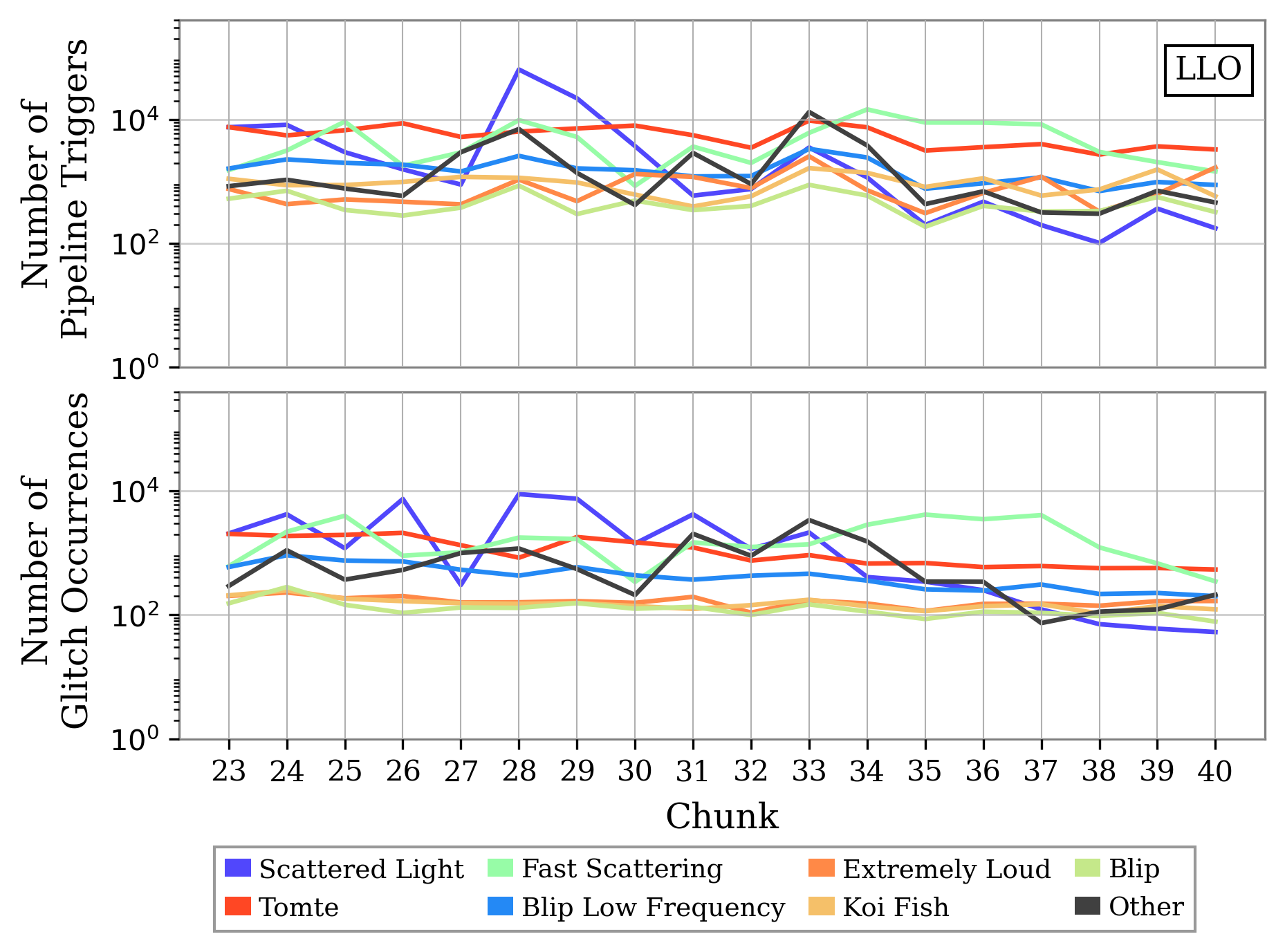}
    \end{subfigure}
    \caption{Number of pipeline triggers caused by specific glitch types (top panels) and number of occurrences of said glitches (bottom panels) at LHO (top) and LLO (bottom) for all chunks in O3b. While trigger rates generally follow glitch occurrence, some classes—like Extremely Loud, Koi Fish, and Tomte—show consistent coupling to the pipeline, while others, such as Scattered Light and Fast Scattering, vary over time. The sharp drop in Scattered Light–induced triggers at LHO after chunk 31 aligns with the commissioning of RC tracking \cite{soni2021reducing}.}
    \label{fig:line-plot}
\end{figure}

Figure \ref{fig:line-plot} shows the number of pipeline triggers caused by each glitch type (top panels) alongside the total number of glitches of each type (bottom panels) for both LIGO Hanford and LIGO Livingston across O3b. 

The number of pipeline triggers caused by each glitch class generally tracks with the number of times that glitch occurs—glitches that appear more frequently tend to produce more pipeline triggers. However, the degree to which each glitch class translates into pipeline triggers also reflects class-specific features and detector conditions.

Some glitch classes show consistent behavior over time. For example, Extremely Loud, Koi Fish, and Tomte glitches at both interferometers exhibit relatively stable relationships between glitch occurrence and pipeline response. Their coupling to the pipeline appears robust across observing chunks, suggesting that these glitches have reliable and repeatable interactions with the template bank.

\begin{figure}[ht]
    \centering
    \includegraphics[width=0.95\linewidth]{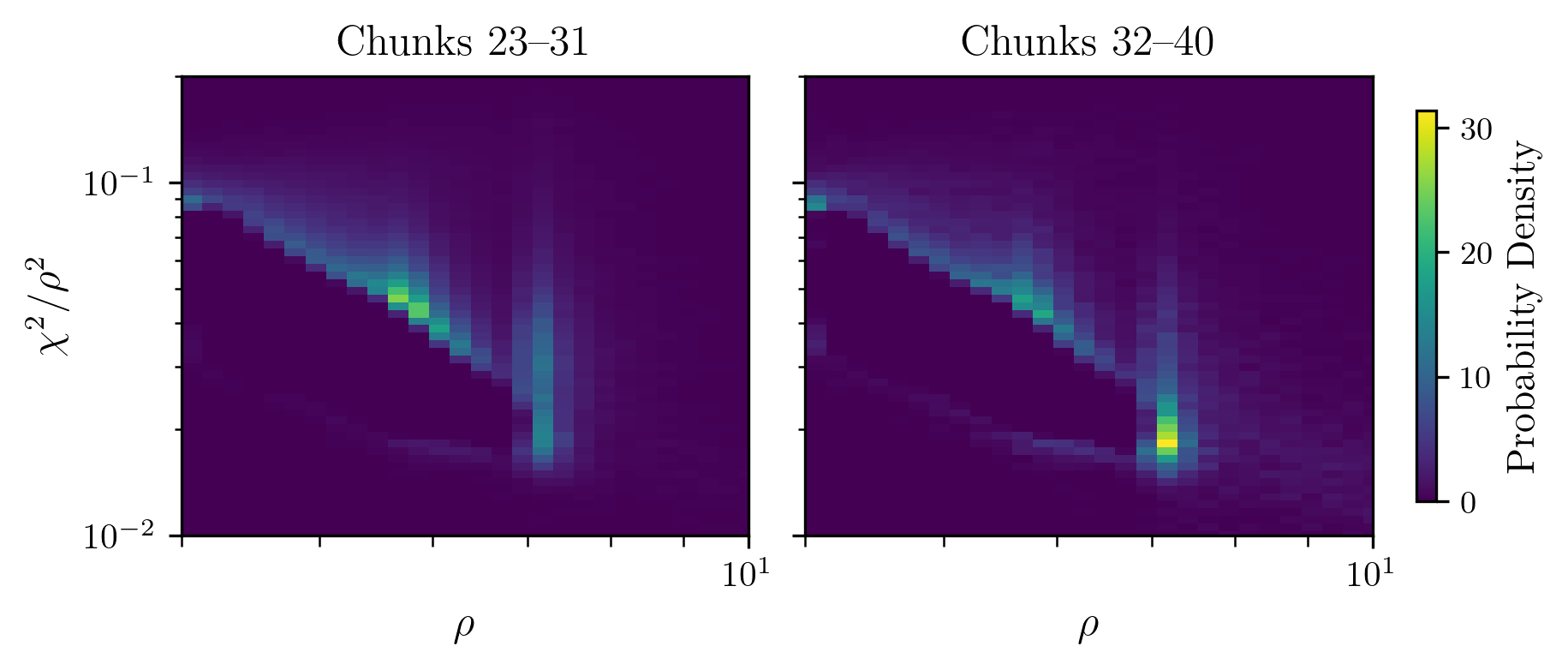}
    \caption{Density distribution of pipeline triggers caused by Scattered Light glitches in the $\rho$ vs.\ $\chi^2/\rho^2$ space, shown separately for chunks 23–31 (left) and chunks 32–40 (right), corresponding to periods before and after the implementation of RC tracking at LHO. There is a distinct shift in the most probable $\rho$ and $\chi^2/\rho^2$ values produced by Scattered Light glitches before vs after RC tracking.}
    \label{fig:rc-tracking-pde}
\end{figure}

In contrast, other glitch types show substantial variability. Both Scattered Light and Fast Scattering fluctuate significantly over time—particularly at LHO. This is consistent with their known dependence on environmental factors such as microseismic motion or ground conditions \cite{LIGO:2024kkz}. A particularly striking example is the sharp drop in Scattered Light–induced pipeline triggers at LHO after chunk 31, which corresponds to the implementation of RC tracking---a commissioning change that successfully mitigated the conditions leading to this glitch type \cite{soni2021reducing}. This commissioning change not only reduced the number of Scattered Light glitches, but also altered how they appeared the search pipeline. Figure~\ref{fig:rc-tracking-pde} compares the $\rho$ vs.\ $\chi^2/\rho^2$ distributions of pipeline triggers caused by Scattered Light glitches before and after the implementation of RC tracking. Prior to the change (chunks 23–31), Scattered Light glitches most frequently produced triggers with $\rho$ values between 5 and 6 and elevated $\chi^2/\rho^2$ values between 0.04 and 0.05. After the change (chunks 32–40), the distribution shifts, with most probable triggers appearing at slightly higher $\rho$ values around 7, but with dramatically lower $\chi^2/\rho^2$ values around 0.001 and lower. The shift in distribution suggests that RC tracking may have successfully eliminated a dominant subclass of Scattered Light glitches, revealing a previously overshadowed population whose behavior now defines the typical pipeline response to this glitch type. This provides a clear example of how changes in detector behavior propagate through to pipeline-level observables, and underscores the utility of pipeline-informed glitch characterization.

Before the implementation of RC tracking, Scattered Light produced the most pipeline triggers in all chunks except chunk 27, where it was second to Koi Fish. This exception is due to a drop in the number of Scattered Light glitches in chunk 27, while the number of Koi Fish glitches remains stable. After RC tracking was implemented, either Koi Fish or Extremely Loud glitches produced the most pipeline triggers across the remaining chunks of O3b. At LLO, there is a greater diversity in the dominant class for each chunk. Two types of stray light---Scattered Light and Fast Scattering---account for the most pipeline triggers in chunks 24, 28, and 29 (Scattered Light), and in chunks 25 and 34–39 (Fast Scattering). Tomte glitches produce the most in chunks 23, 26, 27 and 30-32, and 39-40. Figure \ref{fig:line-plot} shows that the number of Tomte glitches and the pipeline triggers they induce remain relatively consistent across these chunks, whereas there is an increase in the number of Scattered Light glitches and pipeline triggers induced by them for chunks 28 and 29. The origin of Tomte glitches is unknown; however, they were far more common at LLO than LHO throughout O3b.

\subsubsection{Mass 1 vs Mass 2}

\begin{figure}[h]
    \centering
    \includegraphics[width=\textwidth]{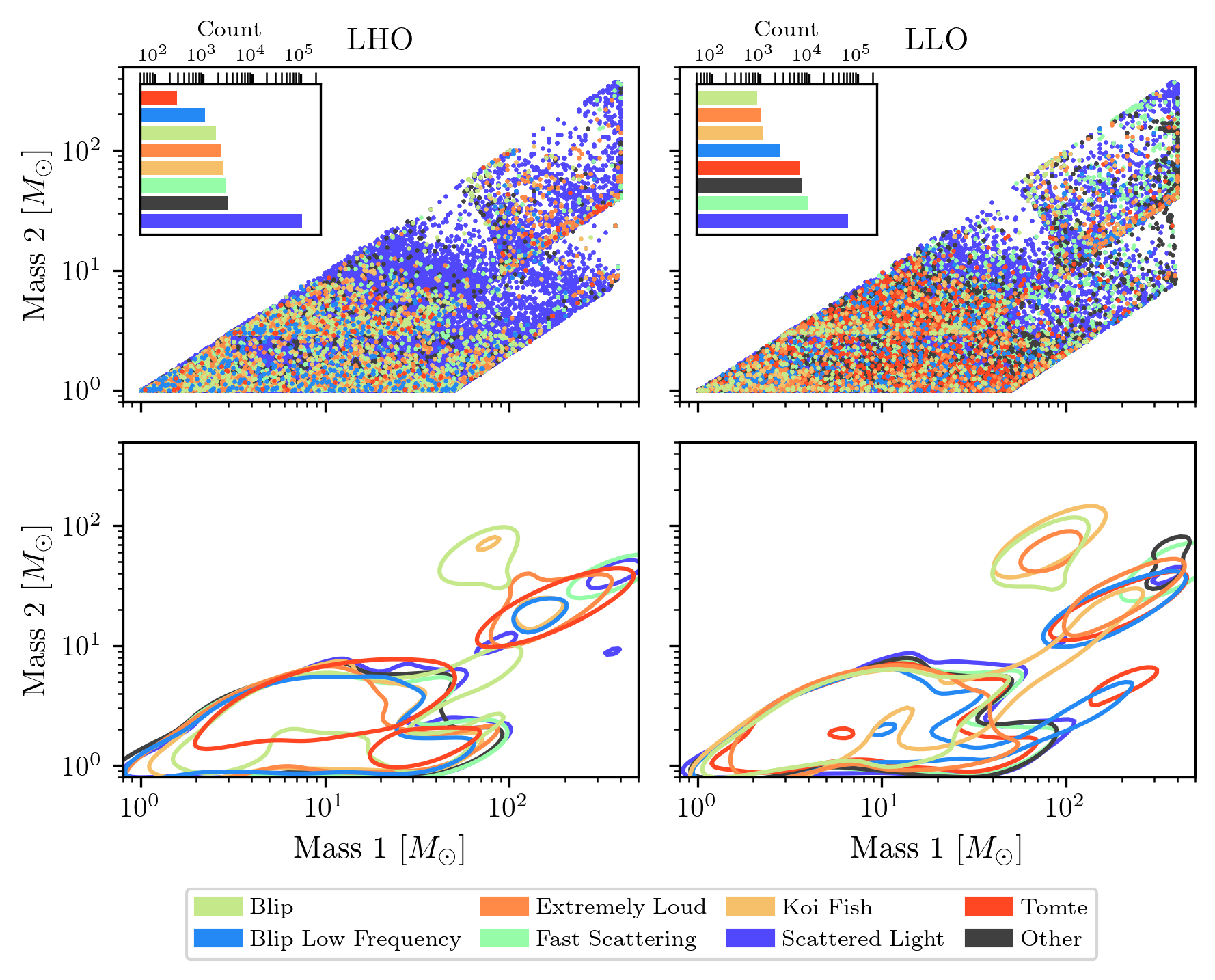}
    \caption{Distribution of Mass 1 vs Mass 2 values for glitch-induced pipeline triggers during chunk 28 of O3b at LHO (left) and LLO (right). On the top row, each point represents a single trigger, colored by the glitch type responsible. The bottom row shows a KDE contour encompassing 90\% of the scatter points to better visualize global trends. Glitch-induced triggers cluster densely in the low-mass, symmetric region associated with BNS templates, reflecting the higher template density in this part of the parameter space. Triggers also extend into regions of high mass ratio and, to a lesser extent, high total mass, indicating that some glitches resonate well with a broad range of CBC template waveforms.}
    \label{fig:m1m2}
\end{figure}

Figure \ref{fig:m1m2} shows the distribution of Mass 1 and Mass 2 for foreground pipeline triggers generated in the 28 chunk of O3 at each interferometer. By convention, Mass 1 is always greater than Mass 2 for each system. Each individual trigger is represented by a scatter point that has been colored according to the type of noise that caused it, and the order of appearance has been determined by the number of triggers caused by each class, with the most populous being plotted first (on the bottom) and then appearing in descending order (this is reflected in the order of the bar plots). Due to the high density of triggers in some regions, individual points may be obscured beneath others. To capture the overall structure of each glitch type, the bottom row shows kernel density estimate (KDE) contours that enclose 90\% of the scatter points, providing a summary of the dominant regions of parameter space occupied by each class.

In both detectors, glitch-induced triggers tend to cluster in the low-mass, symmetric region of the parameter space corresponding to the BNS search region. This is expected to some degree, as the template bank used by the GstLAL pipeline is more densely populated at lower masses to better resolve the longer-duration signals characteristic of BNS systems. Additionally, a notable number of glitch-induced triggers appear at asymmetric mass ratios, particularly at lower total masses. This is consistent with previous work suggesting that glitches can resemble the morphology of high mass ratio inspirals in the time–frequency domain \cite{magee2024mitigatingimpactnoisetransients,Cabero:2019orq}. A smaller number of glitch types also induce triggers at higher total masses, indicating that the influence of transient noise spans a wide range of parameter space.

As shown in Figure \ref{fig:m1m2}, we find that some glitch classes produce pipeline triggers that are more highly concentrated within the BNS region. Blip Low Frequency and Scattered Light are particularly prominent in this regard. Their strong overlap with BNS-like templates reflects their broadband structure in the time–frequency domain. In addition to their complete coverage of the BNS space, these two glitch classes produce triggers in the NSBH and BBH region as well, although there are fewer. This shows that Blip and Scattered Light glitches resonate exceptionally well with the templates in the BNS region, and moderately well with templates in other mass ranges. On the other hand, glitch classes like Extremely Loud, Koi Fish, and Tomte tend to generate triggers that are more dispersed throughout the mass plane, often reaching into the NSBH and BBH regions with higher mass ratios and total masses. This suggests a broader template compatibility with no preferential behavior to any one category for these glitch types, potentially due to their more varied or extended morphology.

Taken together, these results highlight that different glitch classes can leave distinct imprints in the CBC parameter space. The overlap between glitch morphologies and search templates is not uniform—some glitch types show highly localized influence, while others impact a broad region of mass space. This insight demonstrates that some types of noise are more compatible with certain types of CBC waveform templates, and reinforces the idea that search pipelines respond to transient noise in class-specific ways that depend on how the glitch morphologies overlap with the template bank.

\subsection{\texorpdfstring{$\rho$ vs $\chi^2 / \rho^2$}{SNR vs chi squared / SNR squared}}

Another parameter space relevant to the GstLAL search pipeline is the $\rho$ vs $\chi^2 / \rho^2$ space. This space is directly used in the construction of GstLAL’s noise background, making it a natural choice for examining how glitch-induced triggers appear relative to typical background behavior. We use $\chi^2 / \rho^2$ rather than $\chi^2$ alone because this normalized form is familiar within GstLAL and enables better comparison across a range of SNR values. As mentioned in Section \ref{sec:gstlal}, all of the triggers discussed in this work are foreground triggers. 

\begin{figure}[ht]
    \centering
    \includegraphics[width=\textwidth]{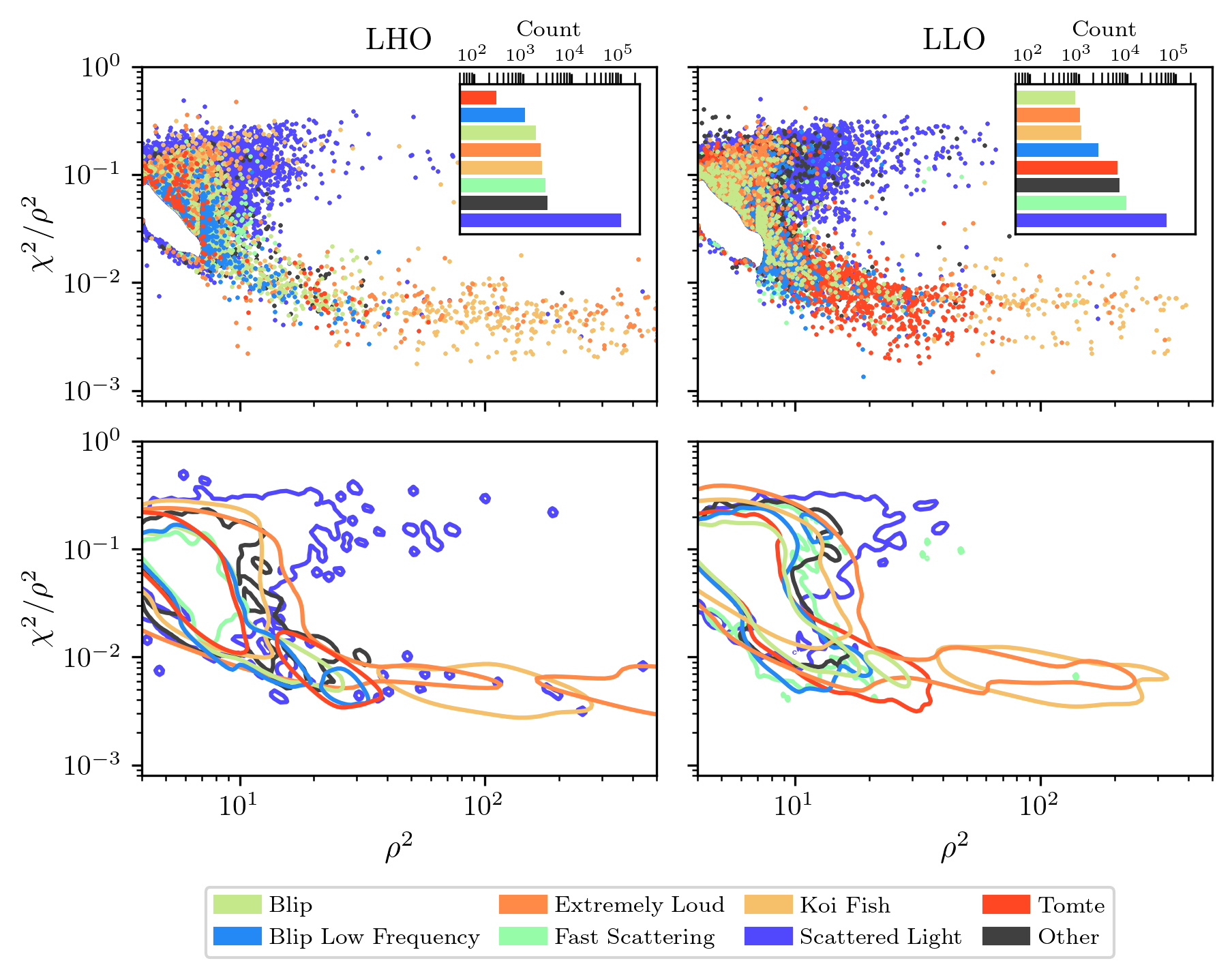}
        \caption{Distribution of $\rho$ vs.\ $\chi^2 / \rho^2$ values for glitch-induced pipeline triggers during chunk 28 of O3b at LHO (left) and LLO (right). On the top row, each point represents a single trigger, colored by the Gravity Spy classification of the glitch that caused it. The bottom row shows a KDE contour encompassing 90\% of the scatter points to better visualize global trends. Distinct glitch types occupy characteristic regions of this space: Scattered Light triggers cluster at low SNRs and high reduced $\chi^2$ values, making them relatively well-contained. In contrast, Extremely Loud and Koi Fish glitches produce triggers with low $\chi^2 / \rho^2$ and high $\rho$, placing them close to the region where genuine signals are expected. Tomte glitches overlap significantly with other glitch types despite having distinct visual morphologies, indicating that visual classification alone may not predict a glitch’s pipeline response.}
    \label{fig:snrchisq}
\end{figure}

Figure \ref{fig:snrchisq}, analogous to the previous mass-space comparison, shows that different glitch types occupy distinct regions of this parameter space. Scattered Light reliably produces triggers with $\chi^2 > 0.1$, nearly all of which have $\rho < 20$, making them relatively well-contained and easier to distinguish from high-significance signals. In contrast, other glitch classes—particularly Extremely Loud and Koi Fish at LHO—produce triggers spanning a wide range of SNRs and consistently low $\chi^2$ values, resulting in increased LRs and placing them uncomfortably close to the region occupied by real astrophysical signals. 

At LLO, glitch behavior follows similar trends: Scattered Light, Extremely Loud, and Koi Fish again dominate broad regions of the $\rho$–$\chi^2$ space. However, Tomte glitches stand out as particularly interesting. Triggers associated with Tomte glitches overlap extensively with regions occupied by several other glitch types, even though their morphologies are quite distinct. This behavior suggests that some unique glitch types produce similarly unique distributions in the pipeline’s parameter space. However, other visually distinct glitches, like Tomte and Koi Fish, may still produce overlapping trigger distributions, indicating that visual morphology alone does not consistently predict how glitches are interpreted by the pipeline. Moreover, the broad parameter space overlap between visually distinct glitch types (like Tomte and Koi Fish) raises the possibility that these glitches may share a common source, even if their morphologies differ.

Our analysis suggests that the distinction between Gravity Spy may not always align with pipeline sensitivity. That is, different glitch categories may induce similar search parameters within the pipeline. Efforts in detector characterization and noise mitigation could therefore add additional context by considering glitch classes based on their influence on search pipelines. Moreover, this work enables us to directly assess how environmental disturbances---such as an increase in Scattered Light glitches due to high microseismic ground motion \cite{LIGO:2024kkz}---are perceived by the search pipeline. By using glitch-induced triggers as a probe of pipeline response, we gain a clearer view of the relationship between environmental factors and pipeline behavior. This approach helps align efforts in detector characterization and noise mitigation with the actual behavior of search pipelines, enabling more mutually informed strategies for improving both.

\subsection{Over-representation of Different Glitch Types}

 Using Equation \ref{eq:overrep}, we calculate the over-representation ratio for several glitch classes at each interferometer. Glitches in the Koi Fish, Extremely Loud, Blip, Blip Low Frequency, Low Frequency Lines, and Tomte classes all exhibit ratios greater than one at both LHO and LLO, indicating that they are disproportionately represented in the pipeline’s trigger population relative to their occurrence rates. In contrast, Scattered Light and Fast Scattering glitches are under-represented at LLO, with ratios below one, and Scattered Light is also under-represented at LHO despite being the most prolific glitch class in terms of absolute trigger count (see Figures \ref{fig:m1m2} amd \ref{fig:line-plot}). This suggests that, while Scattered Light glitches are frequent, each individual glitch may produce relatively few pipeline triggers compared to other classes. This finding highlights that glitch prevalence alone does not necessarily translate to pipeline disruption, and some rarer glitches may have greater influence due to their tendency to consistently generate triggers with parameters closer to genuine gravitational waves.
 
\begin{figure}[ht]
    \centering
    \includegraphics[width=\linewidth]{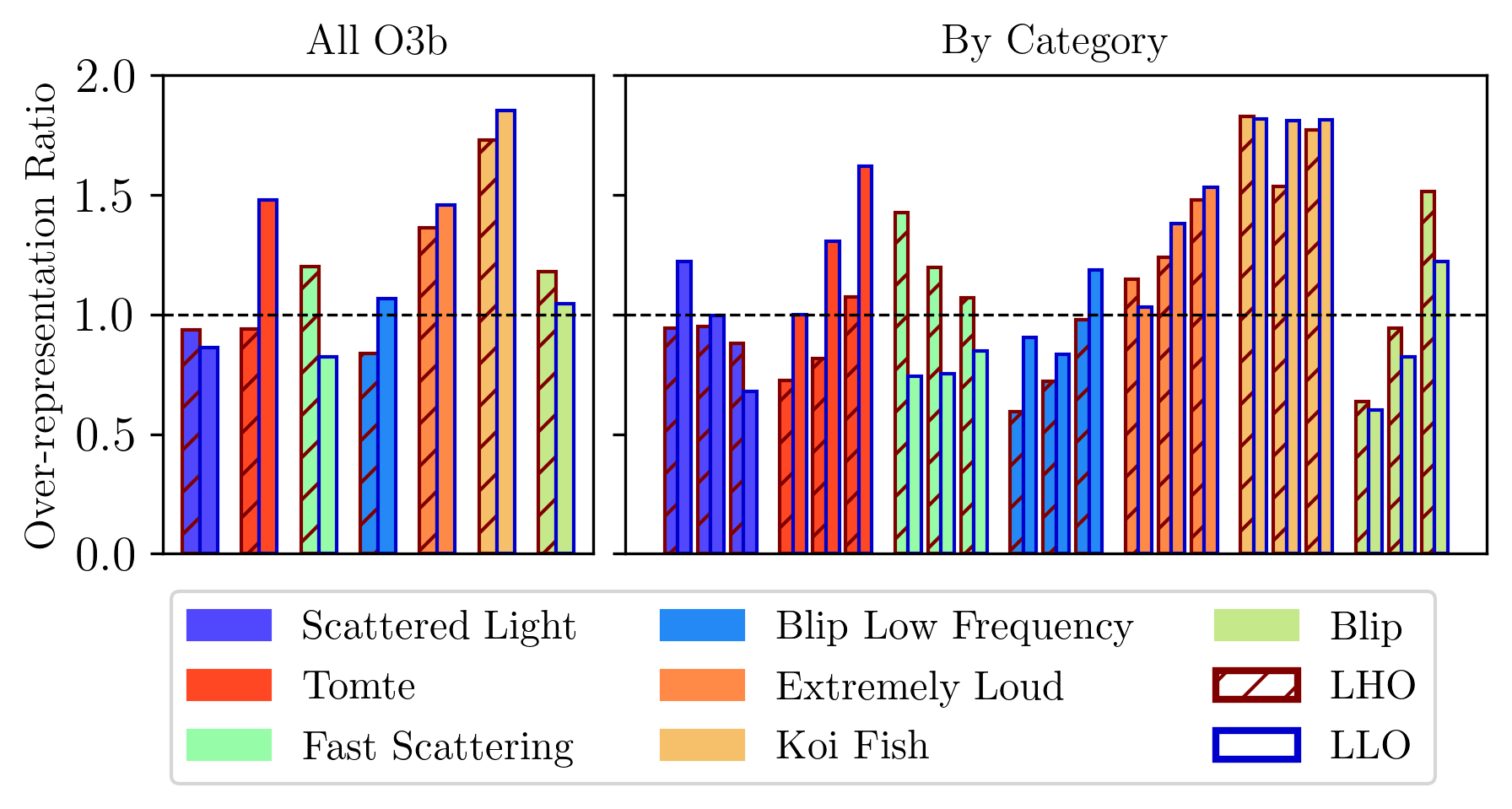}
    \caption{Over-representation ratios by glitch type and search category. The left panel shows the over-representation ratio for each glitch type at both interferometers, calculated as the fraction of triggers caused by a glitch relative to that glitch's overall occurrence. Ratios greater than 1 indicate that a glitch class is disproportionately represented in the population of pipeline triggers. The right panel separates this analysis into search categories---BNS, NSBH, and BBH---shown in order for each glitch type at each interferometer. While most glitch types follow consistent patterns across categories, certain morphologies stand out: Blip glitches are more impactful in the BBH region, while Fast Scattering and Extremely Loud glitches influence BNS searches more significantly.}
    \label{fig:overrep_combo}
\end{figure}

To explore the impact of different glitch types on the search pipeline in more detail, we compute the over-representation ratio within specific search categories---BNS, NSBH, and BBH---for each glitch class. Figure \ref{fig:overrep_combo} presents these category-specific ratios for a selection of prominent glitch types. In most cases, the over-representation values within each category mirror the overall trends shown previously, and the relative behavior between interferometers remains consistent.
However, some deviations emerge. Blip glitches at both detectors deviate from this trend, demonstrating low over-representation in the BNS category and much higher over-representation in the BBH category. This pattern is physically reasonable, as Blip glitches typically appear as short, high-frequency transients in the time–frequency domain---features that overlap more strongly with the shorter-duration, higher-mass BBH templates than with the longer-duration BNS templates. This vulnerability has been well recognized in the gravitational-wave community, and several efforts have targeted the development of methods to mitigate the influence of Blip glitches on BBH searches \cite{Cabero:2019orq,magee2024mitigatingimpactnoisetransients,Zevin:2016qwy,Powell_2018,Nitz:2017lco}. The method and results presented here reinforce those prior observations and help to quantify the scale of the problem in O3b, and provide complementary insight by systematically identifying the specific glitch classes most responsible for contaminating different regions of parameter space.

Building on prior studies that identified the influence of specific glitch types on distinct regions of parameter space, our results show that Fast Scattering glitches at LHO are particularly over-represented in the BNS search space. This is a particularly relevant search space given the scientific return of systems in this mass range, and the prevalence of Fast Scattering is therefore important to understand and address. While the precise mechanisms behind this coupling remain to be fully explored, these findings highlight the value of investigating targeted pipeline improvements tailored to specific search categories, particularly where certain glitch morphologies exert disproportionate influence.

\subsubsection{SNR Distribution}
\begin{figure}[h]
    \centering
    \begin{subfigure}{\textwidth}
        \centering
        \includegraphics[width=\textwidth]{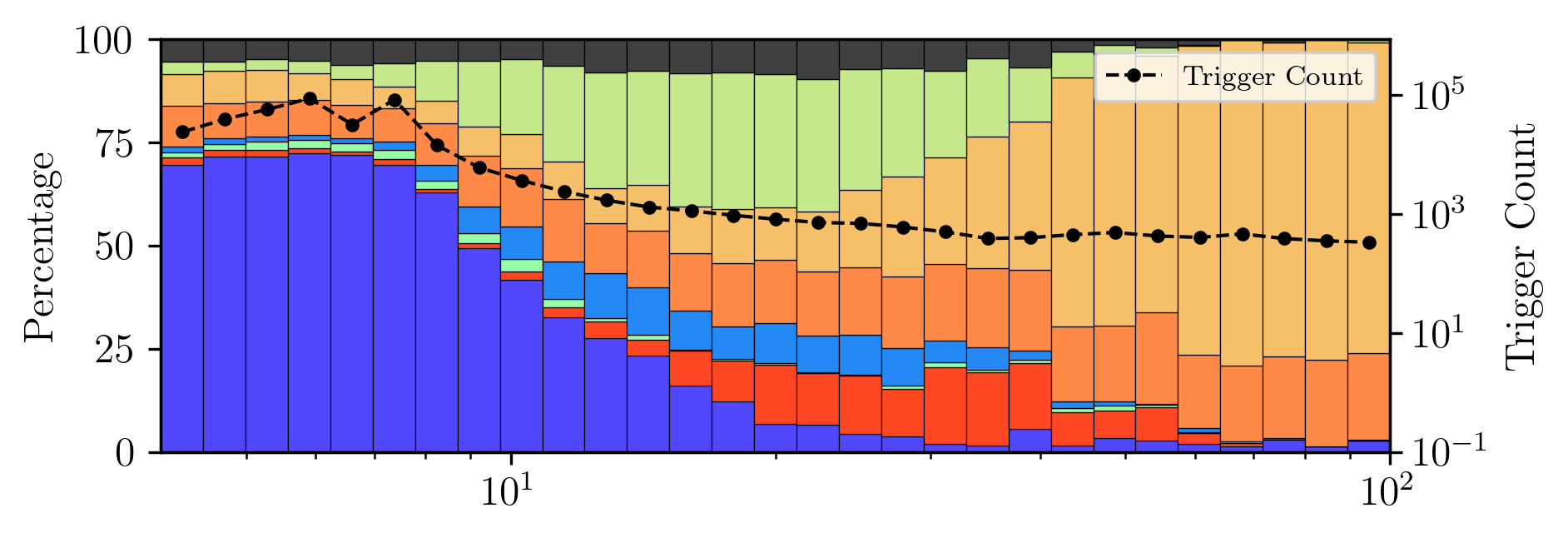}
    \end{subfigure}
        \begin{subfigure}{\textwidth}
        \centering
        \hspace{-0.1cm}
        \includegraphics[width=\textwidth]{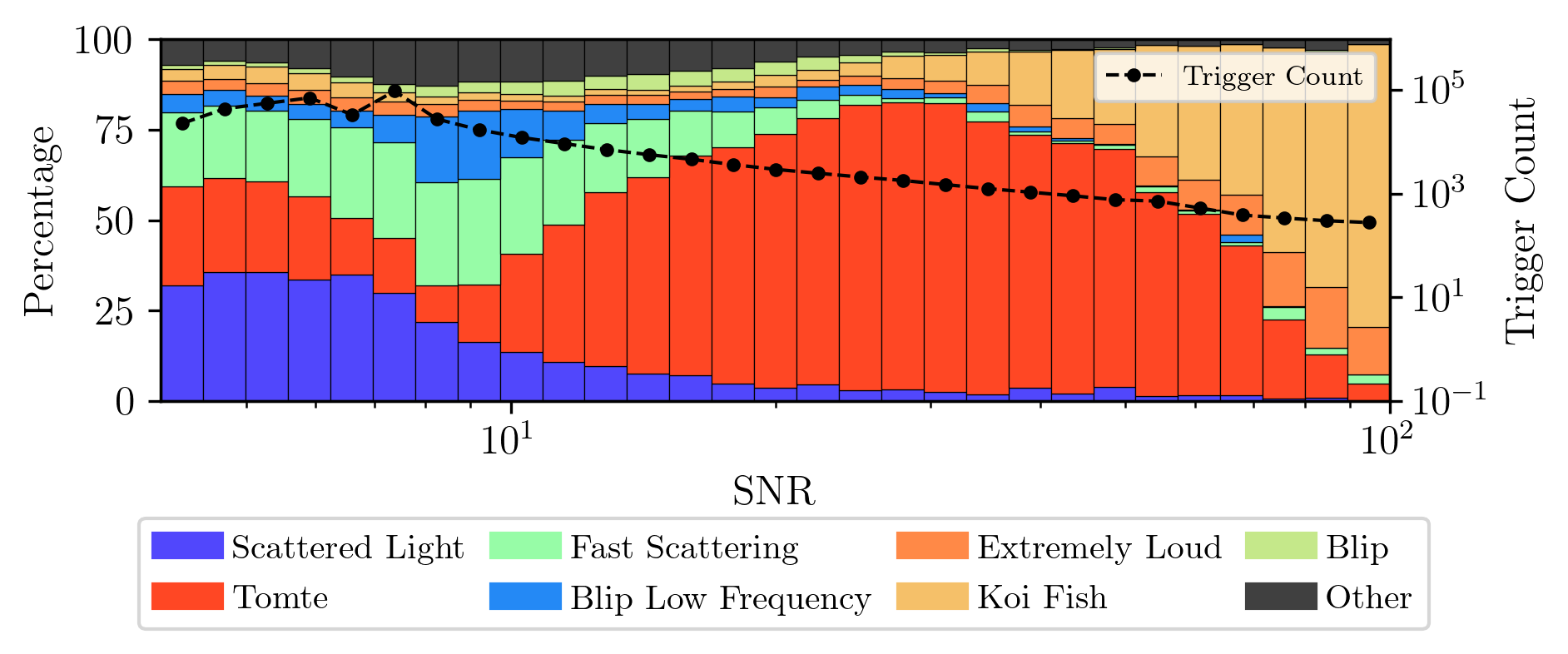}
    \end{subfigure}
    \caption{SNR histogram of all pipeline triggers caused by glitches at each detector over O3b. Each bin has been colored according to the percentage of each glitch type in that bin. Scattered Light makes up the majority of low SNR triggers at LHO, while LLO has a mix of Scattered Light, Tomte, and Fast Scattering dominating the same SNR range. Extremely Loud and Koi Fish glitches produce the overwhelming majority of high SNR triggers at both detectors. At values 10 $\leq$ SNR $\leq$ 100, Tomte is the dominant class at LLO with occasional contributions from Scattered Light and Fast scattering, whereas this SNR range at LHO is occupied by a mix of Scattered Light, Blip, Tomte, Extremely Loud, and Koi Fish glitches.}
    \label{fig:pbp}
\end{figure}

Figure \ref{fig:pbp} shows a histogram of the distribution of the SNRs returned by pipeline triggers produced by glitches at LHO and LLO in O3b. Each bin is colored by the percentage of each glitch type that makes up that bin.

This figure shows the range of pipeline trigger SNR values caused by each individual glitch type at each detector. At both LIGO Hanford and LIGO Livingston, the Extremely Loud and Koi Fish glitch types induce pipeline trigger with a broad range of SNR values, and are almost exclusively responsible for the pipeline triggers with SNR values $\geq 100$. At LIGO Livingston there is a dominant presence of Tomte glitches in the range of $10 \leq$ SNR $\leq 100$, including several bins where Tomtes produce more than $80\%$ of the pipeline triggers. Tomte glitches occur far more frequently at LIGO Livingston than LIGO Hanford, so Tomtes do not dominate the $10 \leq$ SNR $\leq 100$ range at LHO. They do, however, occupy the same range of SNR values that they do at LLO, indicating that, when they do occur, they produce similar pipeline SNR values at LHO, despite happening far less often. The absence of Tomte triggers between $10 \leq$ SNR $\leq 100$ at LHO is filled by a higher presence of Extremely Loud and Koi Fish glitches, and two classes that are not as prevalent at LLO: Blips and Low Frequency Blips. The SNR range occupied by these triggers is within the possible range for a genuine CBC signal. At lower SNR values, Scattered Light produces the most triggers at LHO, and contributes around $20\%$ of the triggers at LLO, joined by triggers produced by Tomtes and Fast Scattering. Overall, this overview gives a clearer picture of the typical SNR ranges produced by different glitch types in the pipeline. It is interesting to see how certain glitches consistently show up in particular SNR bands, such as Tomtes clustering in the mid-range or the dominance of Extremely Loud and Koi Fish glitches at the highest SNRs. These patterns highlight the characteristic ``footprints" different glitches leave in the search pipeline and offer a useful reference point for understanding how noise artifacts map onto SNR.

\subsection{Variation or Consistency Over Time}
\label{sec:variation}

\begin{figure}[ht]
    \centering
    \includegraphics[width=\linewidth]{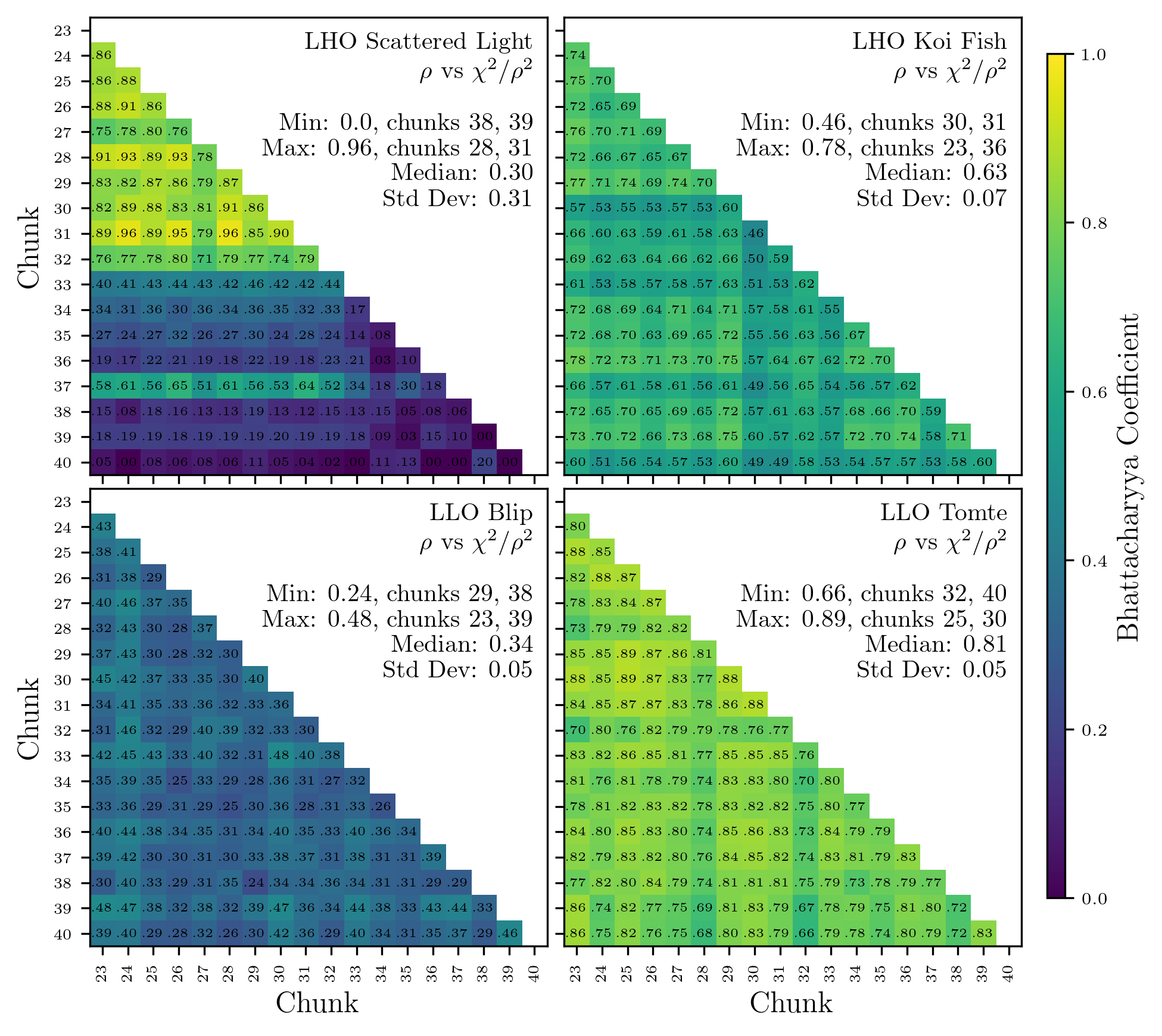}
    \caption{Bhattacharyya Coefficients comparing trigger parameter distributions across time chunks for selected glitch classes at LHO and LLO. Each panel shows the pairwise similarity between chunks for a specific glitch class in a given interferometer, using joint probability distributions in $\rho$ vs. $\chi^2/\rho^2$ parameter spaces. Higher values indicate stronger overlap and thus more consistent trigger parameter distributions over time. Tomte glitches at LLO exhibit strong internal consistency, Scattered Light glitches at LHO show significant variability, reflecting known environmental dependencies. Blip glitches at LLO show moderate consistency, highlighting subtle intra-class variation not evident from morphology alone.}
    \label{fig:bc_matrix}
\end{figure}

Understanding whether a given glitch class produces consistent or variable trigger parameters over time can offer insight into its long-term stability, the pipeline’s response to it, and its potential to be mitigated or modeled. Calculating the Bhattacharyya Coefficient for probability distributions of relevant triggers parameters during a given chunk against the distributions of the same parameters during different chunks allows us to make statements about the similarities of the trigger parameters produced by the same class of glitch over time. This comparison is shown in Figure \ref{fig:bc_matrix} where the Bhattacharyya Coefficient has been calculated using the probability distributions of the $\rho$ vs $\chi^2$/$\rho^2$ spaces at each interferometer.

In the $\rho$ vs $\chi^2$/$\rho^2$ space, Scattered Light glitches at LHO produce triggers that vary in their similarity. This is illustrated by the difference in the minimum and maximum similarity between chunks, with chunks 38 and 39 having no  overlap, and chunks 31 and 28 having greater than 95\% overlap. The standard deviation of 0.31 is indicative of a parameter space that evolves or fluctuates significantly over time. This variability aligns with the changes observed following the implementation of RC tracking, which significantly altered the detector’s Scattered Light behavior. There is a pronounced difference when comparing chunks after the implementation of RC tracking with those before, consistent with the earlier analysis in Figure~\ref{fig:rc-tracking-pde}, where we observed a marked shift in the $\rho$ vs.\ $\chi^2/\rho^2$ distribution following the commissioning change. While chunks prior to RC tracking exhibit relatively high similarity with one another, the chunks after RC tracking show increased variability among themselves. This reinforces our previous observation that RC tracking may have eliminated a dominant, consistent population of Scattered Light glitches, thereby unmasking a more heterogeneous set of glitch behaviors that vary more significantly across time. The observed variability between the chunks after RC tracking is well motivated: the occurrence of Scattered Light is known to be influenced by environmental conditions such as ground motion, which fluctuate over time \cite{LIGO:2024kkz}.
The evolution of the Bhattacharyya Coefficient distribution offers a quantitative reflection of this physical reality. That a known environmental dependency emerges directly from statistical comparisons of trigger parameter distributions reaffirms the method’s ability to reveal physically meaningful trends from the perspective of the pipeline.

Blip glitches, often regarded as a well-defined glitch class, display modest internal consistency. At LHO, their median overlap hovers around 0.34 with a minimum of 0.24 and a maximum of 0.48, suggesting that while these glitches share core characteristics, their manifestation in trigger parameters is not entirely uniform across observing chunks. While Blips may appear visually consistent, this analysis suggests that their representation in the pipeline’s parameter space is more variable. This provides important information that could reflect subtle differences that are not immediately apparent via morphological analysis, offering complementary insights into glitch behavior.

In contrast to Scattered Light and Blip glitches, pipeline trigger parameters produced by Tomte glitches at LLO are consistently similar to one another. The median overlap between any two given distributions is greater than 80\%, indicating that Tomte glitches produce a consistent response in the pipeline. This suggests a persistent underlying mechanism for this glitch type at LLO throughout O3b, and demonstrating that Tomte glitches consistently map to the same region of pipeline parameter space---an insight not previously quantified,and one that could aid in in future efforts to anticipate or flag their presence.

Some glitch classes produce remarkably consistent trigger parameters over time, while others exhibit substantial variability. Both behaviors are informative: consistency across chunks suggests a stable and repeatable interaction with the pipeline, whereas variability may point to evolving distribution of glitch classes, environmental coupling, or changing detector conditions, offering valuable opportunities for identifying and mitigating nonstationary noise sources. The Bhattacharyya Coefficient thus provides a method to interpret not just where glitches reside in parameter space, but how reliably they return there.

\section{Conclusions and Future Applications}
\label{sec:conclusions}
We have introduced PINCH, a method for identifying and analyzing pipeline triggers induced by transient noise using SVM classifiers trained on clean triggers from nominal detector behavior. The trained SVM models assign scores to ambiguous triggers during glitchy periods, allowing us to separate those consistent with random noise from those likely caused by glitches. This enables a novel, pipeline-informed view of how different glitch types manifest in matched-filtering pipelines like GstLAL.

Our results show that different glitch classes affect the pipeline in unique and measurable ways. By analyzing trigger parameters—especially in mass–mass and $\rho$–$\chi^2/\rho^2$ spaces—we find that some glitches (e.g. Tomte, Extremely Loud, Koi Fish) project broadly across search regions and SNR ranges, while others (e.g. Scattered Light) are more narrowly localized. These distinctions reveal how different glitch classes couple to the pipeline’s template bank, shaping where and how they populate the search parameter space.

Our results show that different classes of glitches impact the pipeline in distinct and measurable ways. The mass–mass and $\rho$–$\chi^2/\rho^2$ parameter spaces reveal that some glitches dominate certain search regions. For instance, BNS-like triggers at LLO are most affected by Extremely Loud, Tomte, and Scattered Light glitches—the same classes that heavily impact NSBH and BBH regions. At LHO, the BNS region is dominated by Scattered Light. These trends help identify where in parameter space glitches are most disruptive and provide a roadmap for where sensitivity losses are likely to occur. Some glitch types, such as Tomte and Koi Fish, produce triggers across a wide range of SNR values and mass configurations. Others, like Scattered Light, tend to occupy more confined regions of parameter space, with more predictable and consistent characteristics. The diversity in this behavior indicates that targeted approaches for pipeline development based on the behavior of each glitch class---whether through improved gating, veto development, or adaptive background construction---could provide meaningful improvements. 

The over-representation ratio quantifies how often a glitch type induces pipeline triggers relative to its overall occurrence rate. While Scattered Light is frequent, it is under-represented in the pipeline, suggesting low per-glitch impact. In contrast, Koi Fish glitches, although they occur less frequency, are highly over-represented, posing a greater threat to search fidelity. When considering the impact of different glitch types on specific search areas, Blip glitches are particularly over-represented in the BBH search category, aligning with their short-duration, high-frequency signature, while Fast Scattering glitches at LHO show disproportionate impact in the BNS search space. These results highlight that not all glitches have the same level of impact on gravitational wave searches, as their influence depends not only on how often they occur, but also on how strongly they couple into specific search categories and parameter spaces. Recognizing these patterns can help inform the prioritization of mitigation efforts and support improvements in overall pipeline performance.

Analysis of the Bhattacharyya Coefficient over time reveals additional dimensions of glitch behavior. Some classes, like Tomte, consistently produce triggers with similar parameter distributions, indicating a repeatable and stable pipeline response. Others, such as Fast Scattering and Scattered Light, vary significantly from chunk to chunk. Some of these variations align with known environmental dependencies---such as seasonal changes in ground motion---demonstrating that this technique can reflect physical realities within the detector environment. Additionally, we demonstrate that commissioning changes like RC tracking have an observable effect on the pipeline response to glitch populations. Others are not well understood, motivating further analysis. Understanding which glitches are stable versus volatile can help guide future strategic changes that mitigate individual categories of noise in both detector characterization and pipeline activities.

Analysis of the SNR distribution of pipeline triggers caused by glitches shows that some glitches produce a narrow range of low-SNR triggers, while others extend into high-SNR territory. Tomtes and Extremely Loud glitches, for example, are dominant contributors to high-SNR pipeline triggers at LLO and LHO respectively, while Scattered Light and Low Frequency Lines are confined to lower SNRs. Understanding this SNR structure is crucial for designing future gating strategies and assessing which glitches are most likely to result in false positives.

From a detector characterization perspective, these results offer a new lens through which to evaluate and learn about glitches. In addition to the categorization via visual morphology, this method provides a direct assessment of how each glitch type interacts with the pipeline’s internal logic and parameter space. For example, prominent, common glitches like Scattered Light may be less damaging than rarer but more pipeline-intrusive classes like Extremely Loud. Additionally, the variability or consistency of a glitch class over time can indicate whether the glitch is tied to a stable source or driven by changing environmental conditions. This distinction has practical implications: stable, repeatable glitches may be easier to target with fixed vetoes or gating strategies, while variable glitches may require adaptive or environment-aware solutions. Moreover, we directly observe that commissioning changes like the implementation of RC tracking produce a measurable shift in the parameter distributions of affected glitch populations. Observing a commissioning change through the eyes of the pipeline, we demonstrate the applicability of our approach to connect detector behavior with pipeline response in a way that supports both glitch mitigation and search integrity. By framing glitch impact in terms of how the pipeline ``sees" each class, this work helps bridge the gap between search performance and noise diagnostics, enabling more informed commissioning and mitigation efforts.

Future work will extend this method to O4 data, explore applications across other search pipelines, and investigate real-time implementation. This technique offers a scalable way to diagnose and reduce the impact of non-Gaussian noise in gravitational wave data analysis.

Overall, this work provides a method for identifying pipeline search triggers that are directly caused by transient noise. This distinction allows us to study the interplay between glitches and search pipelines, enabling strategies that are directly informed by the pipeline's perspective. The additional context created by this work will be informative as we search for ways to mitigate non-Gaussian noise and reduce the effect of non-gaussian transients on gravitational wave searches.

\ack{We would like to thank Ryan Magee, Derek Davis, Becca Ewing, Divya Singh, and Patrick Godwin for their comments and suggestions. We would also like to thank the members of the LIGO Detector Characterization Group, the GstLAL team, and the LIGO-Virgo-KAGRA collaboration for their contributions and input. This work was supported by the National Science Foundation under grant numbers PHY-2110509 and PHY-2409740. The material is based upon work supported by NSF's LIGO Laboratory, a major facility fully funded by the National Science Foundation. Additionally, the work utilizes the LIGO computing clusters and data from the Advanced LIGO detectors; the authors are grateful for the computational resources provided by the LIGO Laboratory, supported by National Science Foundation Grants PHY-0757058 and PHY-0823459.}

\section*{References}

\bibliographystyle{iopart-num}
\bibliography{bib}
\end{document}